\documentclass[12pt]{article}
\usepackage{eqsection,subeqnarray,indent,amsfonts,amssymb,amsmath}
\usepackage{bm}    
\usepackage{cite}  
\usepackage{pstricks,pst-node,pst-text,pst-3d,pst-plot}  
\usepackage{type1cm}
\usepackage{graphicx}
\usepackage{setspace}
\usepackage{epsf}
\begin{document}

\newcommand{\be}{\begin{equation}}
\newcommand{\ee}{\end{equation}}
\newcommand{\<}{\langle}
\renewcommand{\>}{\rangle}
\newcommand{\widebar}{\overline}
\def\reff#1{(\protect\ref{#1})}
\def\spose#1{\hbox to 0pt{#1\hss}}
\def\ltapprox{\mathrel{\spose{\lower 3pt\hbox{$\mathchar"218$}}
 \raise 2.0pt\hbox{$\mathchar"13C$}}}
\def\gtapprox{\mathrel{\spose{\lower 3pt\hbox{$\mathchar"218$}}
 \raise 2.0pt\hbox{$\mathchar"13E$}}}
\def\textprime{${}^\prime$}
\def\proof{\par\medskip\noindent{\sc Proof.\ }}
\def\qed{ $\square$ \bigskip}
\def\proofof#1{\bigskip\noindent{\sc Proof of #1.\ }}
\def\half{ {1 \over 2} }
\def\third{ {1 \over 3} }
\def\twothird{ {2 \over 3} }
\def\smfrac#1#2{\textstyle{#1\over #2}}
\def\smhalf{ \smfrac{1}{2} }
\newcommand{\real}{\mathop{\rm Re}\nolimits}
\renewcommand{\Re}{\mathop{\rm Re}\nolimits}
\newcommand{\imag}{\mathop{\rm Im}\nolimits}
\renewcommand{\Im}{\mathop{\rm Im}\nolimits}
\newcommand{\sgn}{\mathop{\rm sgn}\nolimits}
\newcommand{\tr}{\mathop{\rm tr}\nolimits}
\newcommand{\diag}{\mathop{\rm diag}\nolimits}
\newcommand{\Gal}{\mathop{\rm Gal}\nolimits}
\newcommand{\mycup}{\mathop{\cup}}
\newcommand{\Arg}{\mathop{\rm Arg}\nolimits}
\def\hboxscript#1{ {\hbox{\scriptsize\em #1}} }
\def\hboxrm#1{ {\hbox{\scriptsize\rm #1}} }
\def\zhat{ {\widehat{Z}} }
\def\phat{ {\widehat{P}} }
\def\qtilde{ {\widetilde{q}} }
\renewcommand{\emptyset}{\varnothing}

\def\scra{\mathcal{A}}
\def\scrb{\mathcal{B}}
\def\scrc{\mathcal{C}}
\def\scrd{\mathcal{D}}
\def\scrf{\mathcal{F}}
\def\scrg{\mathcal{G}}
\def\scrl{\mathcal{L}}
\def\scro{\mathcal{O}}
\def\scrp{\mathcal{P}}
\def\scrq{\mathcal{Q}}
\def\scrr{\mathcal{R}}
\def\scrs{\mathcal{S}}
\def\scrt{\mathcal{T}}
\def\scrv{\mathcal{V}}
\def\scrz{\mathcal{Z}}

\def\Z{{\mathbb Z}}
\def\R{{\mathbb R}}
\def\C{{\mathbb C}}
\def\Q{{\mathbb Q}}
\def\N{{\mathbb N}}
\def\S{{\mathbb S}}

\def\T{{\mathsf T}}
\def\H{{\mathsf H}}
\def\V{{\mathsf V}}
\def\D{{\mathsf D}}
\def\J{{\mathsf J}}
\def\P{{\mathsf P}}
\def\QQ{{\mathsf Q}}
\def\RR{{\mathsf R}}

\def\bone{{\mathbf 1}}
\def\bv{{\bf v}}
\def\basise{{\bf e}}   
\def\basisf{{\bf f}}   
\def\startv{{\boldsymbol{\alpha}}}   
\def\endv{{\boldsymbol{\omega}}}     

\newtheorem{theorem}{Theorem}[section]
\newtheorem{definition}[theorem]{Definition}
\newtheorem{proposition}[theorem]{Proposition}
\newtheorem{lemma}[theorem]{Lemma}
\newtheorem{corollary}[theorem]{Corollary}
\newtheorem{conjecture}[theorem]{Conjecture}
\newtheorem{question}[theorem]{Question}


\newenvironment{sarray}{
          \textfont0=\scriptfont0
          \scriptfont0=\scriptscriptfont0
          \textfont1=\scriptfont1
          \scriptfont1=\scriptscriptfont1
          \textfont2=\scriptfont2
          \scriptfont2=\scriptscriptfont2
          \textfont3=\scriptfont3
          \scriptfont3=\scriptscriptfont3
        \renewcommand{\arraystretch}{0.7}
        \begin{array}{l}}{\end{array}}

\newenvironment{scarray}{
          \textfont0=\scriptfont0
          \scriptfont0=\scriptscriptfont0
          \textfont1=\scriptfont1
          \scriptfont1=\scriptscriptfont1
          \textfont2=\scriptfont2
          \scriptfont2=\scriptscriptfont2
          \textfont3=\scriptfont3
          \scriptfont3=\scriptscriptfont3
        \renewcommand{\arraystretch}{0.7}
        \begin{array}{c}}{\end{array}}

%
%
\title{Cluster Simulation of the O(N) loop model on the Honeycomb lattice}
\author{Youjin Deng~\footnote{Correspondence should be sent to: 
yd10@nyu.edu} \\ 
Department of Physics, 4 Washington Square Place, \\
New York University, New York NY 10003, USA \\ \\
Wenan Guo \\ Physics Department, Beijing Normal University, \\
Beijing 100875, People's Republic of China \\ \\
Henk W.J. Bl\"ote \\
Faculty of Applied Sciences, Delft University of Technology, \\
P.O. Box 5046, 2600 GA Delft, The Netherlands\\
Lorentz Institute, Leiden University, \\
P.O. Box 9506, 2300 RA Leiden, The Netherlands}
\maketitle

\begin{abstract}
We study the $O(N)$ loop model on the Honeycomb lattice with real 
value $N \geq 1$ by means of a cluster algorithm. The
formulation of the algorithm is based on the equivalence of the 
$O(N)$ loop model and the low-temperature graphical representation
of a $N$-color Ashkin-Teller model on the triangular lattice. 
The latter model with integer $N$ can be simulated by means of 
an embedding Swendsen-Wang-type cluster method. 
By taking into account the symmetry among loops of different 
colors, we develop another version of the Swendsen-Wang-type 
method. This version allows the number of colors $N$ to take 
any real value $N \geq 1$. As an application, we investigate the
$N=1.25, 1.50, 1.75$, and $2$ loop model at criticality. 
The determined values
of various critical exponents are in excellent agreement with 
theoretical predictions. In particular, from quantities 
associated with half of the loops, we determine some 
critical exponents that corresponds to those for the 
tricritical $q=N^2$ Potts model but have not been observed yet. 
Dynamic scaling behavior of the
algorithm is also analyzed. The numerical data strongly suggest 
that our cluster algorithm {\it hardly} 
suffers from critical slowing down. 
\end{abstract}

\section{Introduction}
In Monte Carlo studies of statistical systems undergoing phase transitions,
critical slowing down is one of the prominent problems. Consider a Monte
Carlo algorithm with dynamic exponent $z>0$. In order to generate a
given number of effectively independent samples, one has to spend 
computing effort $\propto L^{d+z}$, where $L$ is the linear sytem size 
and $L^d$ accounts for the volume of the sytem of interest. For the local
Metropolis algorithm for the Potts model, the dynamic exponent is 
around $2.2$. Thus, in two dimensions, the required computing effort 
grows like $L^{4.2}$. Therefore, a central task of 
computational statistical physics is to develop algorithms such
that $z$ vanishes or is significantly suppressed. For a discussion of
Monte Carlo methods, see Ref.~\cite{Binder_84}.

For the Potts model, a significant breakthrough was the 
invention of the Swendsen-Wang cluster method~\cite{Swendsen_87} 
and its single-cluster version--the Wolff cluster method~\cite{Wolff_89}.
For the two- and three-dimensional 
Ising model, the dynamic exponent of the SW algorithm is 
about $0.1$ and $0.45$~\cite{Ossola_2004}, respectively. In comparison with the Metropolis
simulations, the critical slowing down is significantly suppressed.

In addition to the Potts model, another important class of
models in statistical physics is the $O(N)$ model. The $O(N)$ model is 
defined in terms of $N$-component spins on a lattice, with an isotropic
pair coupling of the form $E_{ij}=\epsilon (\vec{S}_i \cdot \vec{S}_j )$,
where $i$ and $j$ are a pair of neighboring lattice sites and $\epsilon$
is a function. A particularly interesting case is the honeycomb 
$O(N)$ model, where function $\epsilon$ is $\epsilon( p) \equiv - 
\log (1 +x p)$, with $x$ a measure of the inverse
temperature. It turns out that the $O(N)$ model has a nice graph 
representation~\cite{Stanley_68,Domany_81,Nienhuis_82,Nienhuis_84}; 
the graph consists of a number of nointeracting and 
nonoverlapping loops on the honeycomb lattice. 

However, in contrast to the Potts model, an efficient cluster algorithm
is still lacking for the $O(N)$ model.  When simulating the $O(N)$ model,
one has to apply a Metropolis-like local
algorithm, except for some special 
cases such as $N=0$ or $N=1$. Even worse is that local updates of 
loop configurations require some global connectivity information.
Therefore, the computing effort grows like $L^{d+z+z'}$ as 
$L$ increases, where the exponent $z'$ accounts for the effective critical
slowing down due to the global-connectivity-checking procedure. The 
value of $z'$ is close to $2$ in two dimensions, unless some complicated
data structure is applied.

Apart from the above computational considerations, developing 
efficient algorithms for the $O(N)$ loop model is also highly 
desirable from physical point of view. This is because, while 
much exact information about the critical properties of the 
$O(N)$ loop model has been accumulated in the past decades,
many open questions still exist; for some recent publications,
see e.g., Refs.~\cite{Chayes_00,Guo_00,Jacobsen_03,Janke_05,Guo_06}.
As a generalization, dilution can be introduced into the $O(N)$ loop 
model and the Potts model. For a sufficient number of diluted sites, 
a different universality class can arise at the so-called tricritical point. 
In contrast to the tricritical 
Potts model, exact results for the tricritical $O(N)$ model are scarce.
Therefore, a high-precision 
numerical study of the $O(N)$ model is of great importance. 

In this work, we solved the long-standing problem of developing an 
efficient algorithm for the $O(N)$ loop model for $N \geq 1$.
We apply the newly developed cluster algorithm, an embedding Swendsen-Wang-type
cluster method, to the honeycomb $O(N)$ loop model without dilution,
for which many exact predictions are available. The numerical data 
confirm the exact predictions by Coulomb gas theory and by 
conformal field theory. Further, the dynamical data imply that, 
for $N>1$, the embedding cluster algorithm {\it hardly} suffers from 
critical slowing down.
%
%
\section{The O(N) loop and the Ashkin-Teller model: exact mapping}

Let us consider a plane graph $\mathcal{G}=(V,E)$ and its dual graph
$\mathcal{G}^*=(V^*,E^*)$. Let $|V|$ and $|E|$ denote the total numbers
of vertices and of edges of $\mathcal{G}$, respectively, and
$|V^*|$ and $|E^*|$ for graph $\mathcal{G}^*$. The dual relation 
between $\mathcal{G}$ and $\mathcal{G}^*$ guarantees $|E|=|E^*|$; 
there is one-to-one correspondence between 
edges of $\mathcal{G}$ and $\mathcal{G}^*$.

On the edges of $\mathcal{G}$, bonds are placed such that 
they form a number of closed paths, or {\em loops}. In addition, it is 
required that these loops cannot share a common 
bond~\footnote{Loops are allowed to intersect in the sense that 
they can share a common vertex.}. The weight of such
a loop configuration is given by $w^{l} N^c$, and 
the partition sum is
\begin{equation}
\mathcal{Z}_\text{Loop} (w, N)  \;=\; \sum\limits_{\text{loops}} 
w^{\ell} N^{c} \; ,
\label{def_Z_loop}
\end{equation}
where the sum is over all satisfied loop configurations. 
Symbol $\ell$ denotes the sum of the lengths of all loops,
and $c$ is the number of loops. In principle,
parameters $w$ and $N$ can be any real or complex numbers. 
In this work, we shall only consider real numbers $w >0$ and $N >0$,
so that we have a probabilistic interpretation.

First, let us further restrict our attention to the case 
where $N \geq 1$ is an integer, and define a $N$-color 
Ashkin-Teller (AT) model~\cite{Ashkin_Teller_43,Fan_72} 
on the dual graph $\mathcal{G}^*$~\footnote{In 
Refs.~\cite{Ashkin_Teller_43,Fan_72}, only $N=2$ was considered. 
Nevertheless, the Ashkin-Teller model with $N \neq 2$ has also 
received considerable research attentions; see e.g., 
Refs.~\cite{Grest_81,Fradkin_84}}.  On every 
vertex $i\in V^*$ of $\mathcal{G}^*$, one simultaneously places 
$N$ independent Ising spins $\sigma^{(m)}$; say they have color 
$m=1,2,\cdots, N$.  On every edge $\langle i , \rangle \in E^*$, 
spins in the same color interact via 
couplings $J_2$, and any two pairs of spins $\sigma^{(m)}$ and $\sigma^{(n)}$
that are in different colors interact via couplings $J_4$.
The latter involves four-spin interactions.
The Hamiltonian reads
\begin{equation}
\mathcal{H}_\text{AT} (J_2,J_4, N)  \;=\; -J_2 \sum_{m=1}^{N} 
\sum_{\langle i , \rangle  \in E*} \sigma^{(m)}_i \sigma^{(m)}_j
-J_4 \sum_{m >n} \sum_{\langle i , \rangle \in E*} 
\sigma^{(m)}_i \sigma^{(m)}_j
\sigma^{(n)}_i \sigma^{(n)}_j \; .
\label{def_H_AT}
\end{equation}
The partition sum is then given by
\begin{equation}
\mathcal{Z}_\text{AT} (J_2,J_4, N)  \;=\; \sum_{\{
\sigma \}} e^{-\mathcal{H}_{\text{AT}} (\sigma) } \; ,
\label{def_Z_AT}
\end{equation}
where the sum is over all spin configurations $\sigma^{(m)}$ with 
$m=1,2,\cdots , N$, as represented by a single symbol $\sigma$. 
Apparently, in the low-temperature region 
$J_2, J_4 \rightarrow \infty$, systems defined by Eq.~(\ref{def_H_AT}) 
have $N^2$ ground states.
The $N=2$ case reduces to the isotropic version of the standard 
Ashkin-Teller model~\cite{Ashkin_Teller_43,Fan_72}.

In this work, we shall concentrate on model~(\ref{def_H_AT}) 
in the infinite-coupling limit: $J_2 \rightarrow \infty$, $J_4 
\rightarrow - \infty$, but $J_2 + (N-1) J_4 =J $ fixed at a finite value. 
In order to see the physical implication of this limit, let us 
consider the spin configurations on the ends of a given edge $e$
of graph $\mathcal{G}^*$. From Eq.~(\ref{def_H_AT}), 
if there are $k$ unequal Ising variables in the same color on 
edge $e$, the Boltzmann weight reads
\begin{eqnarray}
k=0 \; : & \; & \exp [ J_2 N +\frac{1}{2} J_4 N(N-1)] \nonumber \\
k=1 \; : & \; & \exp \{ J_2 (N-2) +\frac{1}{2} J_4 [N(N-1)-2(N-1)] \} \nonumber \\
 & \; & \vdots  \nonumber \\
k\hspace{0.5cm} \;  : & \; & \exp \{ J_2 (N-2 k) 
+\frac{1}{2} J_4 [N(N-1)-2 k (N-1) + k (k-1) ] \} \; .
\label{def_weight}
\end{eqnarray}
Normalizing these weights by that of case $k=0$, one has
Table~\ref{table_weights_AT}.
\begin{table}[htb]
\centering
\begin{tabular}{lc}
\hline\hline
Configuration   & Normalized weight\\
\hline \\[-2mm]
$k=0$    & $1$      \\[2mm]
$k=1$    & $e^{-2 [ J_2 +(N-1) J_4]} \equiv e^{-2J } $      \\[2mm]
$k\geq 2$    & $e^{-2 k J}\cdot e^{k(k-1) J_4} \rightarrow 0 $      \\[2mm]
\hline\hline
\end{tabular}
\caption{\label{table_weights_AT}
Boltzmann weights for the model \protect(\ref{def_H_AT}) 
in the infinite-coupling limit. For any chosen
two pairs of spins $\sigma^{(m)}$ and $\sigma^{(n)}$ with $m \neq n$
we first quote the weight read off from \protect(\ref{def_H_AT}),
and then the normalized weight obtained by making the first one equal to $1$.
We have used $J$ to specify $J_2 +(N-1) J_4$. Notice that the weight 
for $k\geq 2$ vanishes.
}
\end{table}

In words, for any given edge $e \in E^*$, only 
two types of spin configurations are allowed on its ends: 
spins in the same color 
$\sigma^{(m)}_i$ and $\sigma^{(m)}_j $ are all equal, or 
there is {\it at most} one pair of unequal spins in the same color. The 
relative weight of the latter over the former is $e^{-2J}$.
From now on, we shall refer to the AT model whose relative weights
are given in Table~\ref{table_weights_AT} as the infinite-coupling
AT model (IAT).

It turns out that, at least for integer $N$, the loop model 
defined by Eq.~(\ref{def_Z_loop}) can be exactly mapped onto the 
$N$-color IAT model. For this purpose, let us consider the low-temperature 
expansion of the IAT model, which can be performed in a similar way to the 
low-temperature expansion of Onsager's Ising model 
(see e.g., ref.~\cite{Baxter_book}). 
For each color of Ising spin variable $\sigma^{(m)}$, 
we represent those unequal neighboring spins by lines of color $m$ 
on the corresponding edge of the graph $\mathcal{G}$ 
(recall that the IAT model itself is defined on the dual 
graph $\mathcal{G}^*$). 
In words, if two adjacent spins of color $m$ are unequal, 
draw a line of color $m$ on the edge of $\mathcal{G}$; otherwise, 
do nothing. Do this for all pairs of nearest-neighbor spins. 
Apart from effects caused by boundaries~\footnote{One would expect that
boundary effects vanish in the thermodynamic limit.}, 
one obtains an Eulerian subgraph $E'$; i.e., for every vertex, the number 
of lines touching it must be {\em even}. The lines are therefore 
joined together to form polygons (loops); these polygons have color $m$.

Conversely, these polygons divide the plane into 
spin-up and  spin-down domains for Ising variable $\sigma^{(m)}$. For any 
such set of loops, there are just two corresponding spin configurations:
they are related to each other by flipping all spins $\sigma^{(m)}$.

Do this for all colors of Ising variables $\sigma$ (recall that we have used 
symbol $\sigma$ to represent all spin variables $\sigma^{(m)}$ for $m=1,2,
\cdots, N$).  One obtains a configuration containing $N$-color loops.
Any such set of loops corresponds to $N^2$ spin configurations.

\begin{figure}[htb]
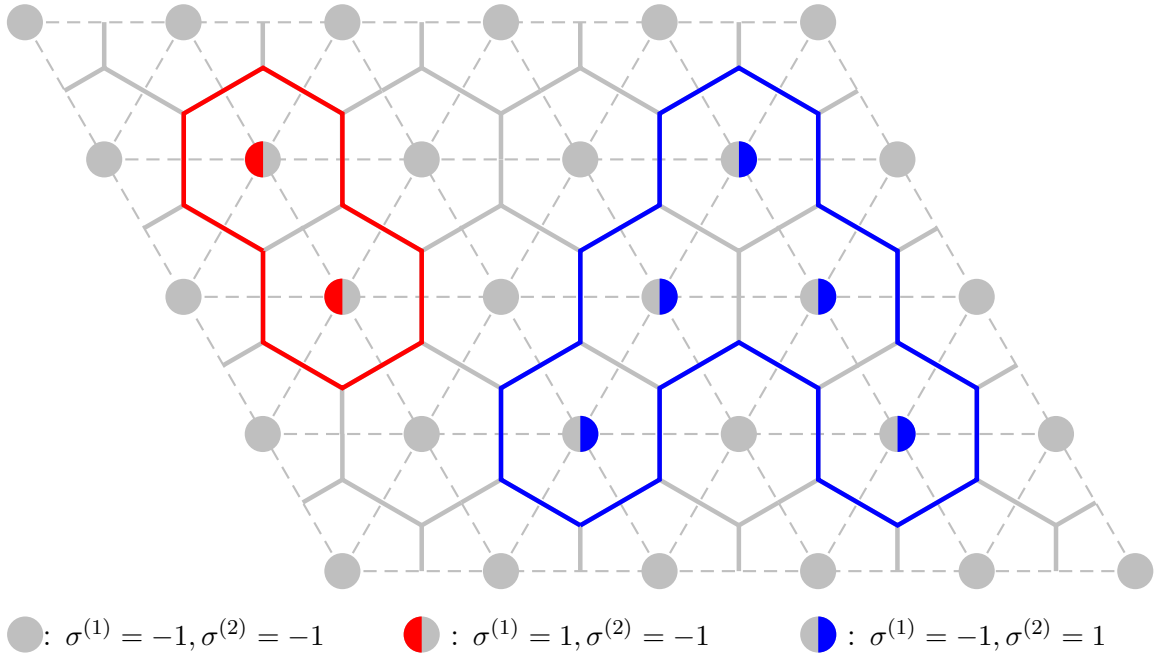

\begin{center}
\psset{xunit=30pt}
\psset{yunit=30pt}
\pspicture(0.5,-1.4)(6,7)
\multirput(-1,1.73205){5}{\psline[linecolor=lightgray,linestyle=dashed]
(0,0)(10,0)}
\multirput(2,0){6}{\psline[linecolor=lightgray,linestyle=dashed]
(0,0)(-4,6.9282)}
\multirput(-1,1.73205){4}{%
 \multirput(2,0){5}{\psline[linecolor=lightgray,linestyle=dashed]
 (0,0)(1,1.73205)}
}
\multirput(-1,1.73205){5}{%
 \multirput(2,0){6}{\pscircle*[linecolor=lightgray](0,0){0.24}}
}
\multirput(-1,1.73205){4}{%
 \multirput(2,0){5}{%
 \psline[linecolor=lightgray,linewidth=0.06](0,1.73205)(0,1.1547)(1,0.57735)(1.5,0.866025)
 \psline[linecolor=lightgray,linewidth=0.06](-0.5,0.866025)(0,1.1547)
 \psline[linecolor=lightgray,linewidth=0.06](1.0,0.57735)(1,0)
}}
\pswedge*[linecolor=blue](3,1.73205){0.24}{-90}{90}
\pswedge*[linecolor=blue](4,3.46415){0.24}{-90}{90}
\pswedge*[linecolor=blue](6,3.46415){0.24}{-90}{90}
\pswedge*[linecolor=blue](7,1.73205){0.24}{-90}{90}
\pswedge*[linecolor=blue](5,5.19615){0.24}{-90}{90}
\pswedge*[linecolor=red](-1,5.19615){0.24}{90}{270}
\pswedge*[linecolor=red](0,3.46415){0.24}{90}{270}
\psline[linecolor=blue,linewidth=0.06](3,0.57735)(4,1.1547)(4,2.3094)(5,2.88675)(6,2.3094)(6,1.1547)(7,0.57735)(8,1.1547)(8,2.3094)(7,2.88675)(7,4.04145)
(6,4.6188)(6,5.7735)(5,6.35085)(4,5.7735)(4,4.6188)(3,4.04145)(3,2.88675)
(2,2.3094)(2,1.1547)(3,0.57735)
\psline[linecolor=red,linewidth=0.06](-1,4.04145)(-1,2.88675)(0,2.3094)
(1,2.88675)(1,4.04145)(0,4.6188)(0,5.7735)
(-1,6.35085)(-2,5.7735)(-2,4.6188)(-1,4.04145)
\pscircle*[linecolor=lightgray](-4,-0.8){0.24}
\rput(-2.0,-0.8){: \small $\sigma^{(1)}=-1, \sigma^{(2)}=-1$}
\pscircle*[linecolor=lightgray](1.0,-0.8){0.24}
\pswedge*[linecolor=red](1.0,-0.8){0.24}{90}{270}
\rput(3.0,-0.8){: \small $\sigma^{(1)}=1, \sigma^{(2)}=-1$}
\pscircle*[linecolor=lightgray](6.0,-0.8){0.24}
\pswedge*[linecolor=blue](6.0,-0.8){0.24}{-90}{90}
\rput(8.0,-0.8){: \small $\sigma^{(1)}=-1, \sigma^{(2)}=1$}
\endpspicture
\end{center}
 \caption{\label{fig_AT_loop}
  A low-temperature graph of the $N=2$ IAT model on a triangular lattice of 
  size $6\times 6$. The edges of the dual lattice -- the honeycomb lattice,
  are shown as solid gray lines. Loops corresponding to Ising variables
  $\sigma^{(1)}$ and $\sigma^{(2)}$ are displayed as thick solid lines 
  in red and in blue, 
  respectively. No loops can intersect on the honeycomb lattice.}
\end{figure}

For the IAT model, the zero weight of configuration
$\sigma^{(m)}_i \neq \sigma^{(m)}_j \,, \sigma^{(n)}_i \neq \sigma^{(n)}_j $
for any $m \neq n$  guarantees that there is at most one pair of
unequal spins in same color on each edge of $\mathcal{G^*}$. 
Therefore, although loops on graph $\mathcal{G}$ can intersect, they 
can never share a common edge. An example of the low-temperature graph 
for the $N=2$ IAT model on the triangular lattice is shown in 
Fig.~\ref{fig_AT_loop}.

Let the energy of a ground state be zero.
Table~\ref{table_weights_AT} tells that, in comparison with a ground state,
a pair of unequal spins must receive an energy penalty $e^{-2J}$, denoted as 
$w$. The partition sum of the IAT model can then be written as
\begin{equation}
\mathcal{Z}_\text{AT} (J, N)  \;=\; N^2 
\sum_{\text{loops}} \sum_{\{\tau_i \}} w^{\ell_i} w^{\ell_j} \cdots \; ,
\label{def_Z_IAT_0}
\end{equation}
where the second sum is over all possible color arrangements for a given 
loop configuration. The factor $N^2$ accounts for the number of ground states.
Note that each loop can randomly take one of the $N$ colors with equal 
probability, and the second sum can then be easily evaluated. This 
leads to
\begin{equation}
\mathcal{Z}_\text{AT} (J, N)  \;=\; N^2 
\sum_{loops} w^{\ell} N^{c} \; ,
\label{def_Z_AT_limit}
\end{equation}
where $\ell$ and $c$ again represent the sum of the lengths of loops 
and the number of loops, respectively. 

Apart from a trivial constant, the partition 
sum~(\ref{def_Z_AT_limit}) is equal to Eq.~(\ref{def_Z_loop}). 
In other words, an exact one-to-$N^2$ mapping has been established between 
the loop model~(\ref{def_Z_loop}) and the $N$-color IAT model defined 
by Table~\ref{table_weights_AT}.  The 
weight $w$ of the line segment in the loop configuration is related to
$J$ by $w =e^{-2J}$.

Note that, if the colors of loops are ignored, the loop representation 
of the IAT model is also a low-temperature graph of the Ising
variable $s \equiv \prod_m \sigma^{(m)}$, the product of all $N$ colors
of variables $\sigma^{(m)}$. This observation is vital in the development of 
cluster algorithms for the $O(N)$ loop model, as shown later.

\section{Cluster simulation of the AT model}

Given the well-known Swendsen-Wang or Wolff cluster algorithm for Onsager's
Ising model, an analogous cluster algorithm is readily 
available for the AT model described 
by Eq.~(\ref{def_H_AT})~\footnote{The word `readily' here might seem misleading, 
since in some recent literature, the very inefficient Metropolis algorithm
was still applied to simulate the standard Ashkin-Teller model in three
dimensions, such as in Ref.~\cite{Musial_2002}.}.
The so-called direct or embedding algorithms can be formulated.
A detailed stuty of the Swendsen-Wang-type algorithms, including 
its dynamical behavior, was carried out by Salas and Sokal~\cite{Salas_97} in 
the context of the $N=2$ AT model on the square lattice (not in the infinite-coupling
limit). In the present work, we shall only consider the embedding version of the 
Swendsen-Wang-type algorithm.

For the AT model~(\ref{def_H_AT}), for a given color of Ising variable
$\sigma^{(m)}$, the effective Hamiltonian, conditioned on the other 
spin configurations $\sigma^{(n)}$ for $n=1,2,m-1,m+1,\cdots, N$, can be
written as
\begin{equation}
\mathcal{H}_\text{eff} (J_2,J_4, N;\sigma^{(m)} )  
\;=\; -J_\text{eff} \sum_{e \in E^*} \sigma^{(m)}_i \sigma^{(m)}_j
\; ,
\label{def_H_eff}
\end{equation}
where the effective nearest-neighbor coupling reads
\begin{equation}
J_\text{eff} = J_2 +J_4 \sum_{n \neq m} \sigma^{(n)}_i \sigma^{(n)}_j \; .
\label{def_J_eff}
\end{equation}
When all pairs of spins $\sigma^{(n)}_i$ and $\sigma^{(n)}_j $ for 
$n \neq m$ are equal, the effective interaction 
$J_\text{eff} = J_2 + (N-1) J_4 \equiv J$ is finite. Otherwise, one
has the coupling $J_\text{eff} =J_2 + (N-1-2 k) J_4 = J -2k J_4 
\rightarrow \infty$. 
Thus, the effective coupling is no longer translation-invariant.
Nevertheless, the effective coupling $J_\text{eff}$ 
is always ferromagnetic. In this case,
when the SW cluster algorithm is applied to update 
the $\sigma^{(m)}$ spins, no frustration
phenomena occur as in cluster simulation of the random Ising model. 
Therefore, it is plausible that the efficiency of such a cluster 
algorithm is not influenced too much by the inhomogeneous effective 
couplings.

A Swendsen-Wang step for updating a given 
Ising variable $\sigma^{(m)}$ can then be written as
\begin{itemize}
\item {\bf Step 1: Place occupied bonds.} Given a spin configuration
$\sigma$, for each edge $e \in E^*$ of $\mathcal{G}^*$ , 
place an occupied bond with probability $p$
\begin{equation}
p = \left\{ \begin{array}{ll}
            1-e^{-2J}  & \mbox{if $\sigma^{(m)}_i = \sigma^{(m)}_j$ and 
                        $\sigma^{(n)}_i = \sigma^{(n)}_j$ for
                         all $n \neq m$ }  \\
            1          & \mbox{if $\sigma^{(m)}_i =  \sigma^{(m)}_j$ and 
                         there is at least one pair
                         $\sigma^{(n)}_i \neq  \sigma^{(n)}_j$} \\
            0          & \mbox{if $\sigma^{(m)}_i \neq \sigma^{(m)}_j$} \; . 
             \end{array}
      \right.
\label{bond_prob}
\end{equation}

\item {\bf Step 2: Construct clusters.} Two vertices 
that are connected through a chain of occupied bonds are said to be in the
same cluster. On the basis of occupied bonds, the whole set of vertices 
$V^*$ of graph $\mathcal{G}^*$ is decomposed into a number of clusters $\mathcal{C}$.

\item {\bf Step 3: Update spins $\sigma^{(m)}$.} For each cluster of $\mathcal{C}$, 
randomly assign the value of all spins on its vertices to be $+1$ or $-1$ 
with equal probability. 
\end{itemize}
This completes a Swendsen-Wang cluster step for spin variable 
$\sigma^{(m)}$. Since all other spin variables $\sigma^{(n)}$ with 
$n \neq m$ are kept fixed, this cluster method is called an embedding
cluster algorithm. We shall refer to this method as the embedding SW
algorithm for the IAT model.

A valid embedding algorithm for the $N$-color AT model must 
involve the update of all colors of Ising spin variables $\sigma$. 
This can be done sequentially or randomly. 

\section{Cluster simulation of the $O(N)$ loop model}

Owing to the exact mapping between the loop and the IAT model, 
for integer $N \geq 1$, the embedding method described in Sec. II is already a 
valid cluster algorithm for the $O(N)$ loop model on graph $\mathcal{G}$. 
In particular, for $N=1$, this cluster method just reduces to the 
standard Swendsen-Wang cluster algorithm for the Ising model on 
graph $\mathcal{G}^*$. The procedure in Sec. II was already 
applied to simulate the $N=2$ IAT model on the honeycomb, square, and 
triangular lattices~\cite{Deng_2006}. In comparison with the Metropolis
method, the critical slowing down is considerably suppressed, but still
observable. At the critical point of the triangular IAT model, which
corresponds both to the critical Baxter-Wu model and to the 
$O(2)$ loop model on the honeycomb lattice, it was found~\cite{Deng_2006} 
that the dynamical exponent $z$ is about $1.18$. In this section,
we shall show that the cluster algorithm in Sec. II can be further 
improved such that it can simulate the loop model with noninteger $N \geq 1$ 
and that critical slowing down {\em hardly} exists.

Let us  consider the Ising spin $s=\prod_{m} \sigma^{\rm (m)} $. 
As mentioned earlier, any
loop configuration of the $O(N)$ loop model on $\mathcal{G}$ is also a 
low-temperature graph of the $s$-spin configurations on $\mathcal{G}^*$.
But information about the colors of loops cannot be encoded in 
the $s$-spin configurations. 

In terms of loop configurations, the simulation of $\sigma^{(m)}$-spin variable 
in the embedding SW procedure in Sec. II is equivalently to
update the loops of color $m$ while keeping all other loops 
of color $n \neq m$ unchanged. 

Our first task is then to rewrite the embedding SW procedure in Sec. II 
by using spin variable $s$ and loop colors as the dynamical variables, 
instead of using $\sigma^{(m)}$
for $m=1,2,\cdots,N$. This leads to the following four steps:
\begin{itemize}
\item {\bf Step 1: Place occupied bonds.} Given a spin configuration $s$
on $\mathcal{G}^*$ and the color information for loops on $\mathcal{G}$, 
for each edge $e \in E^*$, place an occupied bond with probability $p$
\begin{equation}
p = \left\{ \begin{array}{ll}
            1-e^{-2J}  & \mbox{if $s_i = s_j$ } \\
            1          & \mbox{if $s_i \neq s_j$ and $e$ crosses a loop 
                        of color $n \neq m$} \\
            0          & \mbox{if $s_i \neq s_j$ and $e$ crosses a loop 
                        of color $m$} \; . 
             \end{array}
      \right.
\label{bond_prob_1}
\end{equation}
\item {\bf Step 2: Construct clusters} on the basis of occupied bonds.
Note that the $s$ spins on the vertices of a given cluster can have 
different signs.

\item {\bf Step 3: Update spins $s$.} For each cluster,
flip all spins on its vertices with probability $1/2$.

\item {\bf Step 4: Update loops.} Consider the low-temperature graph 
of the $s$ spin configuration, let the colors of the existing loops remain 
unchanged, and assign the color of all newly generated loops to be $m$.

\end{itemize}
Do Steps 1-4 for each of the $N$ colors of loops. Step 4 is redundant 
for the procedure in Sec. II, since the colors of loops are already 
encoded in the spin configuration $\sigma^{(m)}$ with $m =1,2,\cdots,N$.

For any satisfied loop configuration of the $O(N)$ loop model, 
the color of each loop can randomly 
take any value of $N$ colors with equal probability. Therefore, 
instead of keeping all loop colors $n \neq m$ unchanged,
one can randomly reassign the color of each loop in Step 4, 
irrespective of the existing color information. On this basis,
one can reformulate the embedding SW procedure as
\begin{itemize}
\item {\bf Step 1: Construct loops and assign loop colors.} 
Given a spin configuration $s$ on $\mathcal{G}^*$, construct the loop 
configuration on $\mathcal{G}$. For each loop, assign its color to be $1$
with probability $1/N$ and to be $0$ with probability $1-1/N$.
We say loops of color 1 and 0 to be `active' and `inactive', respectively.
\item {\bf Step 2: Update the $s$-spin configuration.} This is done
by the Swendsen-Wang-type procedure, in which the bond-occupation 
probability is
\begin{equation}
p = \left\{ \begin{array}{ll}
            1-e^{-2J}  & \mbox{if $s_i = s_j$ } \\
            1          & \mbox{if $s_i \neq s_j$ and $e$ crosses a loop
                        of color 0 } \\
            0          & \mbox{if $s_i \neq s_j$ and $e$ crosses a loop
                        of color 1} \; .
             \end{array}
      \right.
\label{bond_prob_2}
\end{equation}
\end{itemize}
A cluster simulation of the $O(N)$ loop model can then be 
achieved by repeating the above  two steps. Note that $N$ is {\em 
no longer} required to be an integer; it is sufficient to demand
that $N \geq 1$, since $1/N$ should not be bigger than 1.

In summary, in comparison with the embedding SW procedure described in 
Sec. II, the last version of the algorithm has two prominent 
features. First, $N $ is no longer required to be an integer. 
Second, after a completion of the update of Ising spin variable, 
the loop colors are randomly reassigned, irrespective of the existing 
colors. This is possible because all the $N$ colors of spin variables
$\sigma^{(m)}$ ($m=1,2,\cdots,N$) are symmetric to each other. In other
words, the symmetry among the $N$-color spin variables is 
used in the last version of the embedding cluster algorithm. 
If this symmetry is broken in some way, e.g., one can let
couplings $J_2$ depend on spin colors, the last version is then not
applicable. But a slightly modified version of the procedure in Sec. II
still works.

Because of usage of the symmetry among the $N$-color spin variables,
one would expect that modes of the critical slowing down, associated
with this symmetry, are significantly suppressed in the last version 
of the embedding SW algorithm. Thus, one expects that it is 
more efficient than the procedure in Sec. II. This will be numerically
confirmed later.

\noindent
{\bf Remark:} For a plane graph $\mathcal{G}$ of degrees
$d > 3$, the loop assignment based on the $s$ spin configuration
is not unique, since loops are allowed to share a common vertex.
Nevertheless, this should not affect the validity of the present embedding
SW cluster algorithm.

We conclude this section by pointing out that our embedding SW cluster
algorithm for the $O(N)$ loop model is of a similar spirit to
the Chayes-Machta algorithm~\cite{Chayes_98} for the $q$-state 
Potts model with real $q \geq 1$.

\section{The Honeycomb $O(N)$ loop model}

The Honeycomb lattice is of degree 3, and thus the loops of the $O(N)$
loop model never intersect. In this case, it has been known that 
the loop model can be exactly mapped on an $O(N)$ spin model with 
partition sum~\cite{Domany_81,Nienhuis_82,Nienhuis_84}
\begin{equation}
\mathcal{Z} (w, N) = \int \prod_k d \, \vec{S}_k \prod_{\langle i j
\rangle \in E} (1 +w \vec{S}_i \cdot \vec{S}_j ) \; ,
\label{def_ON}
\end{equation}
where $E$ is the edge set of the honeycomb lattice. Here, $\vec{S}$ 
represents a $N$-dimensional unit 
vector -- namely $|\vec{S}| =1$~\footnote{In Ref.~\cite{Nienhuis_82},
the $O(N)$ spin model is defined such that $|\vec{S}| =\sqrt{N}$ and 
the factor $w$ is replaced by $\lambda / N$}.
This model has been
under intensive investigation in the past two decades, 
and much exact information is available.
For $N <2$, the model undergoes a second-order phase transition; for 
$N=2$, the transition is of infinite order--i.e., the so-called 
Kasteleyn-Thouless (KT) transition; for $N >2$, it displays a lattice-gas-type
transition. The renormalization exponents for critical 
$O(N)$ loop systems for $N \leq 2$ is a function of $N$. For $N >2$,
it is expected that the transition is in the same universality class of
Baxter's hard-hexagon lattice gas model on the triangular lattice (an exact mapping
between the $N \rightarrow \infty$ loop model and the hard-hexagon lattice gas 
can be established).

The critical frontier for $N \leq 2$ is exactly located 
at~\cite{Nienhuis_82,Nienhuis_84}
\begin{equation}
w_c (N) = \left(2+ \sqrt{2-N} \right)^{-1/2} \; .
\label{critical_point}
\end{equation}

It was observed~\cite{Nienhuis_82,Nienhuis_84} 
that a {\em critical} $O(N)$ model corresponds with a 
{\em tricritical} $q =N^2$-state Potts model. From exact calculations,
conformal field theory~\cite{Cardy_87}, 
and Coulomb gas theory~\cite{Nienhuis_84}, exact values of 
renormalization exponents can be obtained for the tricritical Potts model as
\begin{eqnarray}
y_{t1} & = & 3-\frac{6}{g} \nonumber \\
y_{t2} & = & 4-\frac{16}{g} \nonumber \\
y_{t3} & = & 5-\frac{30}{g} \nonumber \\
y_{h1} & = & 2-\frac{(6-g)(g-2)}{8g} \nonumber \\
y_{h2} & = & 2-\frac{(10-g)(g+2)}{8g} \; ,
\label{critical_exponent}
\end{eqnarray}
where the Coulomb-gas couplings $g$ is related to $q$ as 
\begin{equation}
g = 4 +\frac{2 \arccos \left( \frac{q-2}{2} \right)}{ \pi}  \; .
\label{relation_gq}
\end{equation}
The symbols $y_{t1}$, $y_{t2}$, and $y_{t3}$ represent exponents of 
the leading, subleading, and next-subleading thermal scaling 
fields, respectively, and $y_{h1}$ and $y_{h2}$ are for the magnetic 
scaling fields. Therefore, exact values of critical exponents for 
the $O(N)$ model described by Eq.~(\ref{def_ON}) as 
a function of $N$ are also known.

\section{Simulation and Sampled quantities}

We applied the embedding SW cluster algorithm in Sec. III 
to simulate the $O(N)$ loop model on the honeycomb lattice; the 
Ising variable $s$ is defined on the vertices of the dual 
lattice--i.e., the triangular lattice.
The sampled quantities can be classified into three types; they
are associated with loop configurations, Ising spin variables, and
distributions of clusters $\mathcal{C}$ formed in the SW step.

The first type includes the total lengths of loops and of the number of loops,
as normalized by the volume of the system $V$ and denoted by $\rho_l$ and 
$\rho_c$, respectively. Here, the volume $V$ is the total number of 
vertices on the corresponding triangular lattice. The second-moment
fluctuations of $\rho_l$ are also sampled as $C = V (\langle 
\rho_l ^2 \rangle - \langle \rho_l \rangle^2 )$, where symbol
$\langle \; \rangle$ represents the statistical average.

Taking into account the mapping between the loop model and the IAT
model, we also measured both the total lengths and the number of 
`active' loops, denoted by $\rho_{la}$ and $\rho_{ca}$; 
the fluctuations of $\rho_{la}$, denoted as $C_a$, were also measured. The 
subscript '$ a$' means `active'.

The second type of quantities concern the Ising variables associated with
loop configurations. A natural quantity to sample is the 
magnetic susceptibility $\chi = 
V \langle m^2 \rangle $, where $m$ is the magnetization density for 
spin variable $s$. After each embedding SW step, we also 
reconstructed Ising spins $\sigma^{(1)}$ from these active loops. 
The associated susceptibility $\chi_a$ was then sampled. In addition, 
we consider the size distribution of domains enclosed by those 
active loops. These domains can be obtained by placing occupied 
bonds between all equal nearest-neighbor $\sigma^{(1)}$ Ising spins.
They are named the Ising clusters in the standard Ising model.
We measured the second moment of these Ising clusters 
\begin{equation}
S_{2a} = \frac{1}{V^2} \sum_i  a_i ^2 \; ,
\end{equation}
where $a_i$ is the number of $\sigma^{(1)}$ spins in the $i$th domain.

The last type of quantities contains information for the embedding SW
simulations. They include the number of occupied bonds per 
spin $n_b$, the number of clusters per spin $n_c$, and the second and
fourth moment of cluster sizes 
\begin{equation}
W_k = \frac{1}{V^k} \sum_i  c_i ^k \; ,
\label{def_wk}
\end{equation}
where $k=2$ or $4$, and $c_i$ is the size of the $i$th cluster formed
in the embedding SW simulations.

In the conventional SW simulation of the Ising model (or more generally 
of the Potts model), quantities $W_2$ and $W_4$ are exactly related to the
second and the fourth moment of the magnetization density. The bond-number
density is also exactly related to the energy density in the Ising model.

On the basis of the above sampled quantities, we sampled ratios
\begin{equation}
Q = \frac{\langle \chi \rangle^2}{\langle \chi^2 \rangle} \; , \hspace{5mm}
Q_a = \frac{\langle \chi^{(1)} \rangle^2}{\langle \chi^{(1)} \; ^2 
\rangle} \; , \hspace{5mm} \mbox{and} \hspace{5mm}
Q_w =  \frac{\langle W_2 \rangle^2}{\langle 3 W_2 ^2 - 2 W_4 \rangle}  \; .
\label{def_ratio}
\end{equation}
At criticality, these amplitudes ratios are dimensionless and universal.
They are known to be very useful in locating critical points 
in Monte Carlo studies of statistical systems undergoing phase transitions.
For the SW simulation of the Ising model, quantity 
$Q_w$ reduces to $Q$ defined on the basis of magnetic susceptibility.

\subsection{Test of the algorithm}
For a test of the algorithm, we performed simulations for $N=1.5$ 
and $N=1.9$. The system sizes took $12$ values in 
range $ 8 \leq L \leq 256$, where $L$ is the linear system size of the
triangular lattice. Periodic boundary conditions were applied. 
Measurements were taken after every Swendsen-Wang step.

\begin{figure}
\begin{center}
\leavevmode
\epsfxsize 10.0cm
\epsfbox{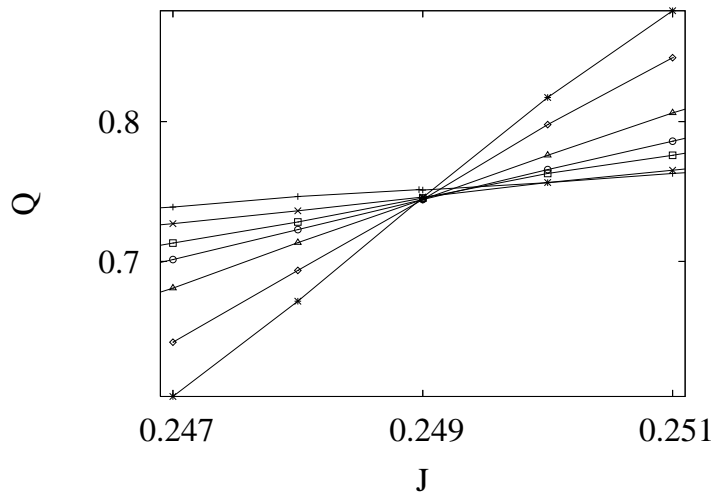}
\end{center}
\caption{Binder ratio $Q$ for $N=1.5$. The data points $+$, $\times$,
$\Box$, $\bigcirc$, $\bigtriangleup$, $\Diamond$,
and $\ast$  represent
system sizes $L=8, 16, 32, 48, 80, 160$, and $256$, respectively.
The error bars of the data are smaller than the point sizes.
The lines, which simply connect data points for each $L$, are
just for illustration. }
\label{fig_Qm1_5}
\end{figure}

For $N=1.5$, Eq.~(\ref{critical_point}) yields the critical point at 
$J_c = - (\ln w_c)/2 = 0.24897\cdots $. Simulations were performed
in range $ 0.245 \leq J \leq 0.253$. About $6 \times 10^6$ samples were
taken for each size $L$. Parts of the $Q$ data are shown in Fig.~\ref{fig_Qm1_5}.
The intersections of the $Q$ data for different system sizes rapidly
converge to the expected value $J_c =0.24897\cdots $. This implies the 
validity of the embedding cluster algorithm for the $O(N)$ loop model
with noninteger $N$.

According the least-squares criterion, we fitted the $Q$ data by
\begin{equation}
Q(J,L) = Q_c+ \sum_{k=1}^{m} (J-J_c)^k L^{k y_t} +
b_i L^{y_i} +b_1 L^{y_1} + \cdots \; ,
\label{fit_q1}
\end{equation}
where $m>1$ is an integer. The terms with amplitudes $b_i$ and $b_1$
describe finite-size corrections. The term with exponent $y_i$ is supposed
to arise from the leading irrelevant scaling field. Thus, the value of 
$y_i$ is given by $y_{t3}$ in Eq.~(\ref{critical_exponent}).
For $N=1.5$, Eq.~(\ref{critical_exponent}) yields $y_i \equiv y_{t3} =
-1.097296 $. Other finite-size corrections can arise from various sources,
such as from the subleading irrelevant scaling field, from the 
analytical part of the free energy $f$, and from the second derivative 
of $f$ with respect to the leading irrelevant field. It is not 
{\em a prior}
clear how many or which such correction terms can be observed in the
numerical data. Thus, one should make various fits by taking into 
account all possible sources of finite-size corrections. Final 
estimates of interested quantities
are then obtained by comparing the results of various fits.
We let exponent $y_t$ and $y_i$ to be fitted by the numerical data,
and fixed $y_1$ at $-2$ for simplicity. All the $Q$ data in range 
$ 8 \leq L \leq 256$ can be successfully described by Eq.~(\ref{fit_q1}).
The results are $y_t =0.749 (4)$, $y_i =-1.4(3)$, $J_c=0.24896(2)$,
and $Q_c=0.7429 (8)$. 
The value of $y_t$ is consistent with the exact 
value $y_{t2}=0.7481\cdots$ 
in Eq.~(\ref{critical_exponent}), and that of $y_i$ agrees 
with $-1.097\cdots$. The estimated critical point is also in excellent 
agreement with the exact prediction $0.24897\cdots$.

\begin{figure}
\begin{center}
\leavevmode
\epsfxsize 10.0cm
\epsfbox{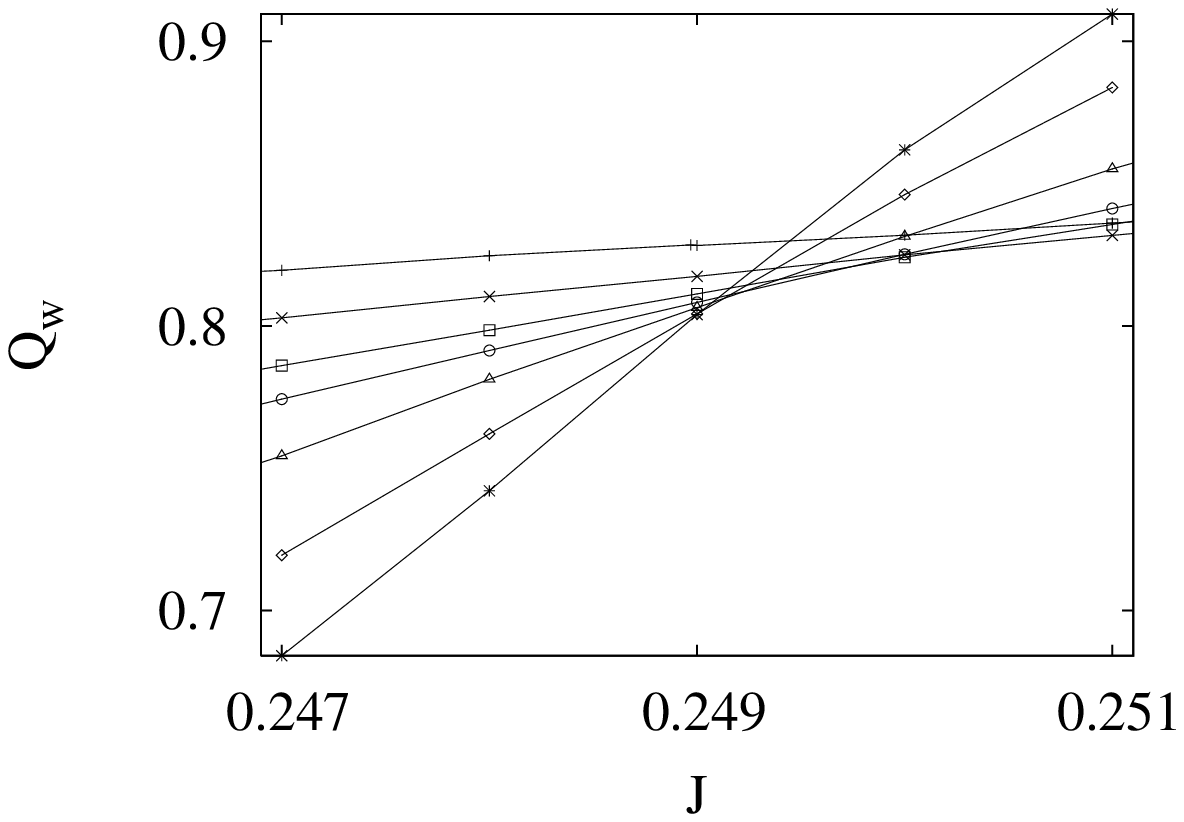}
\end{center}
\caption{Binder ratio $Q_w$ for $N=1.5$. The data points $+$, $\times$,
$\Box$, $\bigcirc$, $\bigtriangleup$, $\Diamond$,
and $\ast$  represent
system sizes $L=8, 16, 32, 48, 80, 160$, and $256$, respectively.\
he error bars of the data are smaller than the point sizes.
The lines, which simply connect data points for each $L$, are
just for illustration purpose. }
\label{fig_Qw1_5}
\end{figure}

Next, we plotted parts of the $Q_w$ data in Fig.~\ref{fig_Qw1_5}. The 
intersections of the $Q_w$ data reflect the percolation threshold of 
clusters formed in the embedding SW procedure. In cluster simulations 
of statistical systems, the efficiency can be reflected by the average 
size of the formed clusters. The efficiency will be limited if 
the average size is either too small or too big: little change of 
configurations is made for the former, and large amount of effort 
has to be carried out for the latter (probably it also accompanies 
by little change). Ideally, the percolation threshold of clusters
formed in Monte Carlo simulations should coincide or be very close to the 
phase transition of statistical systems. This is indeed the case
in the present embedding SW algorithm for the $O(N)$ loop model,
as reflected by Fig.~\ref{fig_Qw1_5}.

The fits of the $Q_w(L)$ data by Eq.~(\ref{fit_q1}) yields $y_t=
0.756(6)$, $y_i=-1.2 (3)$, and $J_c=0.24900(3)$. The estimated 
percolation threshold is consistent with the thermal critical point
$J_c=0.24897\cdots$.

\begin{figure}
\begin{center}
\leavevmode
\epsfxsize 10.0cm
\epsfbox{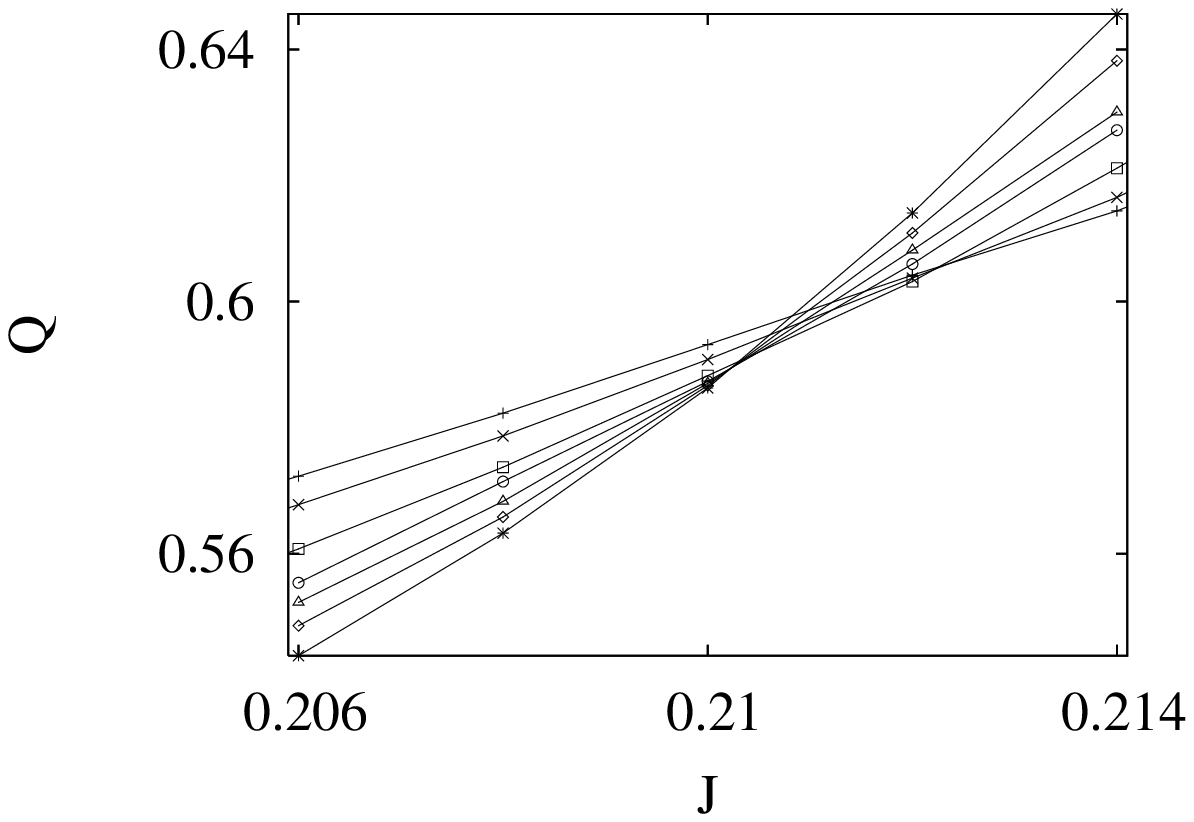}
\end{center}
\caption{Binder ratio $Q$ for $N=1.9$. The data points $+$, $\times$,
$\Box$, $\bigcirc$, $\bigtriangleup$, $\Diamond$,
and $\ast$  represent
system sizes $L=8, 16, 32, 48, 80, 160$, and $256$, respectively.\
he error bars of the data are smaller than the point sizes.
The lines, which simply connect data points for each $L$, are
just for illustration purpose. }
\label{fig_Q1_9}
\end{figure}

We also simulated the $N=1.9$ loop model. 
Equation~(\ref{critical_point}) predicts
the critical point at $J_c = 0.20998\cdots$. Simulations were performed
in range $0.206 \leq J \leq 0.214$, and system sizes took $12$ values
in range $8 \leq L \leq 256$. Parts of the $Q$ data are 
shown in Fig.~\ref{fig_Q1_9}. As expected, the intersections of the 
$Q$ data for different sizes converge rapidly to $J_c=0.20998\cdots$.
The fits of the $Q$ data by Eq.~(\ref{fit_q1}) yields $y_t =
0.373 (6) $, $y_i =-1.2 (2)$, $J_c=0.21000(8) \approx 0.20998\cdots $, 
and $Q_c=0.586(1)$.  The value of $y_t$ is more or less consistent with 
that of $y_{t2}=0.367\cdots$. But that of $y_i$ does not agree with 
that of $y_{t3}=-1.81\cdots$. In this case, we expect that 
the dominant finite-size corrections are not described by 
exponent $y_{t3}$. 

A plausible scenario for the dominant finite-size corrections 
might be the following. As mentioned earlier, the mapping
between the $O(N)$ model on graph $\mathcal{G}$ and the IAT model
on the dual graph $\mathcal{G}^*$ is only exact up to boundary 
effects. Even though these boundary effects are expected to vanish
in the thermodynamic limit $L \rightarrow \infty$, they cannot be 
neglected for finite $L$. According to a simple argument, these
boundary effects vanish as a function of $1/L$. Indeed, exponent $-1$ is
consistent with the numerical values of $y_i$ for $N=1.5$ and $N=1.9$.

\section{Simulation at criticality}

For $N=1.25, 1.5, 1.75$, and $2$, we performed extensive simulations right 
at the exactly predicted critical points given 
by Eq.~(\ref{critical_point}). 
Periodic boundary conditions 
were applied, and the system sizes took $18$ values in range $ 4 \leq L \leq 
1024$. Samples were taken after every embedding SW step. 
The number of samples is $10^7$ for $L \leq 256$, and $2 \times 10 ^6$ 
for $L > 256$ (Wenan and Henk: Monte Carlo data are not complete yet). 

For later convenience, we list in Table.~\ref{table_exponent_ON} the exact
values of critical exponents for $N=1,1.25,1.50, 1.75$,and $2$, 
as predicted by Eq.~(\ref{critical_exponent}); also included are the 
exact values of critical points given by Eq.~(\ref{critical_point}).
\begin{table}[htb]
\centering
\begin{tabular}{r|lllll|r}
\hline\hline
  $N$              & $y_{t1}$        & $y_{t2}$        & $y_{t3}$  
& $y_{h1}$         & $y_{h2}$        & $J_c$  \\
\hline \\[-2mm]
  $1$              & $1.875$         & $1$             & $-0.625$    
& $1.9479\cdots$   & $1.1979\cdots$  & $0.274653\cdots$ \\[2mm]
  $1.25$           & $1.8327\cdots$  & $0.8873\cdots$  & $-0.8361\cdots$
& $1.9343\cdots$   & $1.1562\cdots$  & $0.263231\cdots$  \\[2mm]
  $1.50$           & $1.7805\cdots$  & $0.7481\cdots$  & $-1.0972\cdots$
& $1.9198\cdots$   & $1.1069\cdots$  & $0.248970\cdots$ \\[2mm]
  $1.75$           & $1.7078\cdots$  & $0.5542\cdots$  & $-1.4607\cdots$
& $1.9034\cdots$   & $1.0420\cdots$  & $0.229072\cdots$ \\[2mm]
  $2.00$           & $1.5$           & $0 $            & $-2.5$         
& $1.875$          & $0.875$         & $0.173286\cdots$ \\[2mm]
\hline\hline
\end{tabular}
\caption{\label{table_exponent_ON}
Exact values of critical points and exponents for $N=1,1.25,1.75,2$, as predicted 
by Eqs.~(\ref{critical_point}) and (\ref{critical_exponent}).}
\end{table}

\subsection{$N=2$: KT point}
The critical $N=2$ loop model on the honeycomb lattice is a very 
special case. It can be mapped both onto the standard $N=2$ IAT model
on the triangular lattice and the Baxter-Wu model with three-spin 
interactions on the triangular lattice. The latter has been exactly
solved by various methods, and is believed
to be in the same universality class as the $4$-state Potts model. Further,
it is known that the $N=2$ loop model undergoes a KT transition, which 
is an infinite-order phase transition.

\begin{figure}
\begin{center}
\leavevmode
\epsfxsize 10.0cm
\epsfbox{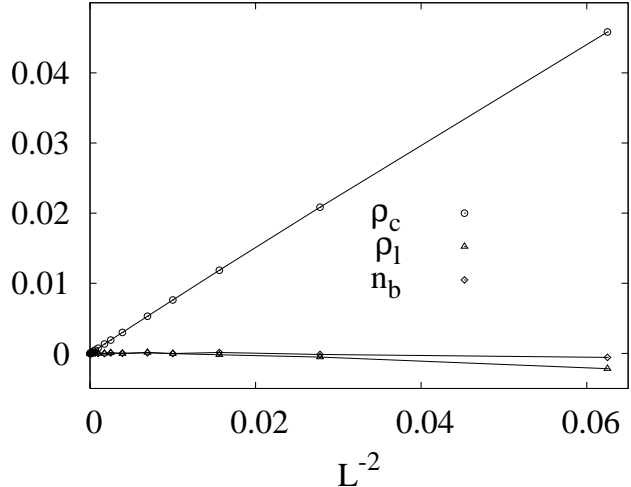}
\end{center}
\caption{Quantities $\rho_l$, $\rho_c$ , and $n_b$ for $N=2$. In order
to display the finite-size dependence for different quantities, 
the analytical-background 
contributions in these quantities have been subtracted. Namely, 
the data in the vertical axis are $\rho_l (L) -\rho_{l0}$, 
$\rho_c (L) -\rho_{c0}$, and $n_b (L) -n_{b0}$, where $\rho_{l0}=
1.114835(4)$, $\rho_{c0}=0.0574189(7)$, and $n_{b0}=1.10956(2)$
were obtained from fits by Eq.~(\ref{fit_energy}). The exponent $-2$
of $L$ in the horizontal axis is $y_{t2}-2$.}
\label{fig_E2_00a}
\end{figure}

{\bf Energy-like quantities.} The energy-like quantities include the 
density of the length of loops $\rho_l$ and of the loop 
number $\rho_c$. The exact information for the critical properties of
the $O(2)$ model tells one that, at criticality, the finite-size 
behavior of $\rho_l$ and $\rho_c$ is governed by the subleading 
thermal exponent $y_{t2}$ in Eq.~(\ref{critical_exponent}).
From the analysis of Monte Carlo data, as shown later, it turns 
out that the finite-size scaling behavior of
bond-occupation density $n_b$, a quantity associated with 
the embedding SW step, is also described by $y_{t2}$.
Instead of being listed in tables, the $\rho_l$ and $\rho_c$
data are shown in Fig.~\ref{fig_E2_00a}; the vertical axis shows
the data for $\rho_l (L)- \rho_{l0}$, $\rho_c (L)- \rho_{c0}$, and
$n_b (L)- n_{b0}$, where constants $\rho_{l0}$, $\rho_{c0}$, and 
$n_{b0}$ were obtained from the numerical fits.

We made a least-squares fit of the Monte Carlo data 
for energy-density-like quantities to the general ansatz 
\begin{equation}
E(L) = E_0 + L^{y_t -d} (a +b_0 L^{-1} +b_i L^{y_i} + \cdots ) \; .
\label{fit_energy}
\end{equation}
The symbols $E$ and $E_0$ should be replaced by specific quantities in
the fits. For instance, depending on which quantity 
is going to be analyzed, $E$ can be $\rho_l$, $\rho_c$, and 
$n_b$, and $E_0$ can be $\rho_{l0}$, $\rho_{c0}$, and $n_{b0}$.

In Eq.~(\ref{fit_energy}), the term with $b_0$ describes boundary
effects arising from the mapping between the Ising-spin configuration and 
the loop representation. The term with $b_i$ arises from the least
irrelevant scaling field. However, the value of $y_i$ is not very clear. 
According to conformal field theory, if an operator with exponent 
$y$ exists, a sequence of exponents $y-1$, $y-2$, $\cdots$, can 
in principle also exist. Therefore, $y_i$ can be $y_{t3}$ or
$y_{t2}-1$ etc. In the analysis of our Monte Carlo data, we have tried
to take into account all possible sources of finite-size corrections.

The fits yield $y_t=0.1(2)$ and $\rho_{l0}=1.114835(4)$ for 
the $\rho_l(L)$ data, $y_t=0.004(8)$ and $\rho_{c0}=0.057418(7)$ 
for the $\rho_c(L)$ data, and $y_t=-0.5(10)$ 
and $n_{b0}=1.10956(2)$ for the $n_b (L)$ data. The values of $y_t$
are all in good agreement with the prediction $y_{t2}=0$.

Within the sampled quantities, the density of the lengths and of the
number of active loops, $\rho_{la}$ and $\rho_{ci}$, are also 
energy-like quantities. Since the color of each loop can randomly 
take one of $N$ colors, the statistical mean value 
of $\rho_{la} (L)$ is just $\rho_c (L) /N$ for any size $L$; the 
same applies to quantity $\rho_c$. Therefore, we do not need to 
analyze them.

\begin{figure}
\begin{center}
\leavevmode
\epsfxsize 10.0cm
\epsfbox{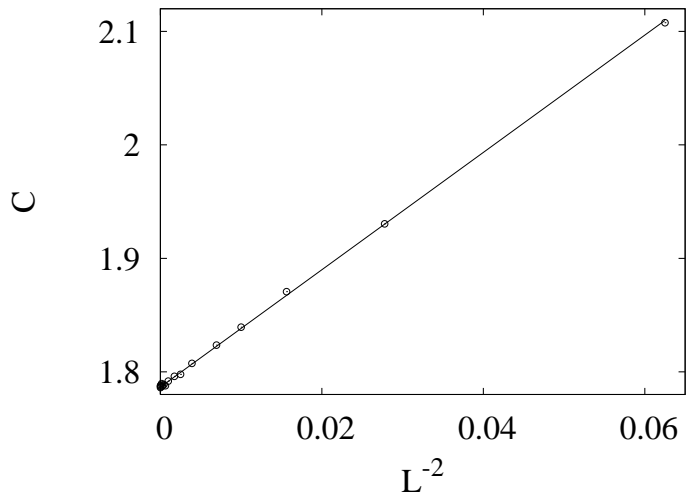}
\end{center}
\caption{Specific-heat-like quantity $C$ for $N=2$. 
The exponent $-2$ of $L$ in the horizontal axis is equal
to $2 y_{t2}-2$.}
\label{fig_C2_00a}
\end{figure}

\begin{figure}
\begin{center}
\leavevmode
\epsfxsize 10.0cm
\epsfbox{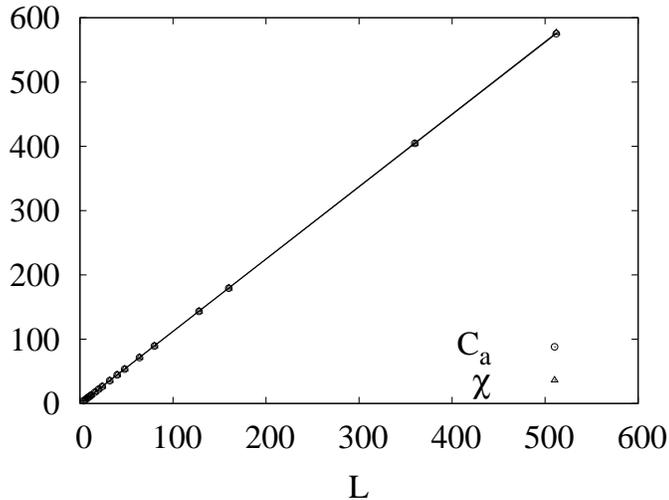}
\end{center}
\caption{Quantities $C_a$ and $\chi$ for $N=2$.
The exponent $1$ of $L$ in the horizontal axis is equal
to $2 y_{t1}-2$. In this scale, the $C_a(L)$ and $\chi(L)$
data collapse for each $L$. }
\label{fig_Ci2_00a}
\end{figure}

{\bf Specific-heat-like quantities.} Quantities $C$ and $C_a$ account
for the fluctuations of $\rho_l$ and $\rho_{la}$, respectively. 
The $C$ and $C_a$ data are respectively shown 
in Figs.~\ref{fig_C2_00a} and \ref{fig_Ci2_00a}. These suggest that
the finite-size behavior of $C$ is governed by exponent $2y_{t2}-2 =-2 $,
while that of $C_a$ is governed by $2y_{t1}-2 =1 $. 

An observation
is that the behavior of the magnetic susceptibility $\chi$, for 
the $s = \prod \sigma^{(m)}$ for $m=1,2,\cdots,N$, is also described by 
exponent $2y_{t1}-2 =1 $. Actually, for the present case $N=2$, because
of the symmetry between two Ising variables $\sigma^{(1)}$ and 
$\sigma^{(2)}$ for the IAT model, it can be shown that, in the thermodynamic
limit $L \rightarrow \infty$, $C_a (L)$ and $\chi (L)$ are equivalent 
to each other (Wenan Henk: this statement has not been carefully checked).

The data for specific-heat-like quantities were fitted by
\begin{equation}
\mathcal{C} (L) = \mathcal{C}_0 + L^{2y_t -d} 
(a +b_0 L^{-1} +b_i L^{y_i} + \cdots ) \; ,
\label{fit_specific_heat}
\end{equation}
where symbols $\mathcal{C}$ and $\mathcal{C}_0$ should be replaced
by specific quantities in the fits.

The fits of the $C (L)$ data yield $y_t = -0.01(3) = y_{t2}$ and 
$C_0 =1.7867(8)$, and those of $C_a$ give  $y_t = 1.5002(4) 
= y_{t1}$ and $C_{a0}=-0.46(3)$.

\begin{figure}
\begin{center}
\leavevmode
\epsfxsize 10.0cm
\epsfbox{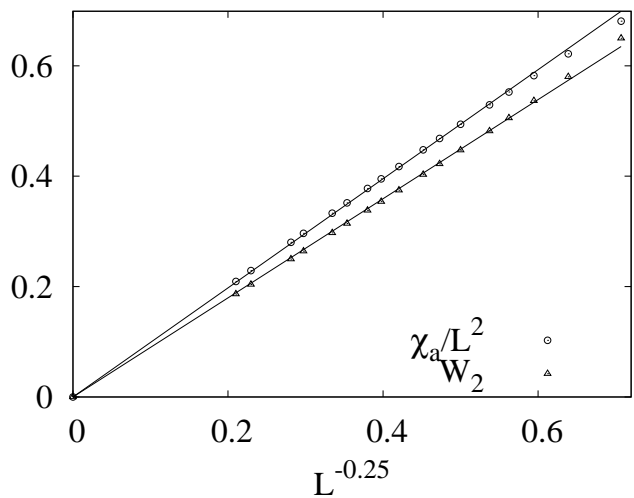}
\end{center}
\caption{Susceptibility-like quantities $\chi_a/L^2$ and $W_2$ for $N=2$.
The exponent $-1/8$ of $L$ in the horizontal axis is equal
to $2 y_{h1}-4$.} 
\label{fig_Mi2_00a}
\end{figure}

\begin{figure}
\begin{center}
\leavevmode
\epsfxsize 10.0cm
\epsfbox{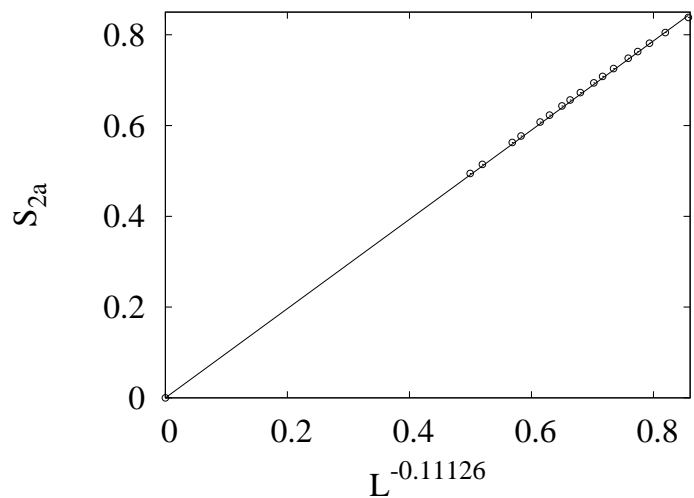}
\end{center}
\caption{Quantity $S_{2a}$ for $N=2$.
The exponent $-0.11126$ of $L$ in the horizontal axis is 
obtained from the fits.}
\label{fig_Si2_00a}
\end{figure}

{\bf Susceptibility-like quantities.} Quantity $\chi$ for the $s$ 
Ising variable has been discussed earlier. The fits of the $\chi(L)$
data also yield $y_t =1.5002(4) =y_{t1}$. Other susceptibility-like
quantities include $\chi_a $ for spin $\sigma^{(1)}$, 
$ S_{2a}$ for the Ising clusters of spin variable $\sigma^{(1)}$, and
$W_2$ for the clusters formed in the embedding SW step. The data
of these quantities are shown in Figs.~\ref{fig_Mi2_00a} and 
\ref{fig_Si2_00a}.

The data for susceptibility-like quantities were fitted by
\begin{equation}
\mathcal{K} (L) = \mathcal{K}_0 + L^{2y_h -d}
(a +b_0 L^{-1} +b_i L^{y_i} + \cdots ) \; ,
\label{fit_chi}
\end{equation}
where symbols $\mathcal{K}$ and $\mathcal{K}_0$ should be replaced
by specific quantities in the fits. Exponent $y_h$ is left 
to be fitted by numerical data.

The fits of the $\chi_a (L)$ and $W_2 (L) $ data by Eq.~(\ref{fit_chi}) 
yield $y_h = 1.8751(1)$ and 
$y_h = 1.8751(2)$, respectively. Both results are in good agreement
with the exact value of $y_{h1}=15/8=1.875$, as given in 
Table~\ref{table_exponent_ON}. This implies that 
the percolation threshold of the clusters formed in the embedding SW
step coincides with the thermal critical point $J_c (N=2) =\ln 2/4$.

The fit of $S_{2a}$ yield $y_h=1.9444(1)$. There is no prediction
of the exact value for this exponent.

\begin{figure}
\begin{center}
\leavevmode
\epsfxsize 10.0cm
\epsfbox{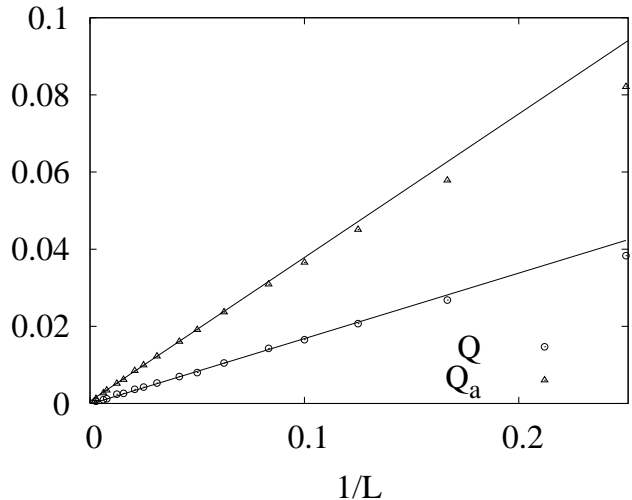}
\end{center}
\caption{Ratios $Q$ and $Q_a$ for $N=2$.
The exponent $-1$ of $L$ in the horizontal axis is
can be explained as $y_{t2}-1$; it is also in good agreement 
with the fitting results for $Q$ and $Q_a$.}
\label{fig_Q2_00a}
\end{figure}

{\bf Binder ratios.} In the Monte Carlo simulations, we also samped 
dimensionless amplitude ratios $Q$ and $Q_a$, defined by 
Eq.~(\ref{def_ratio}). The $Q$ and $Q_a$ data are shown in 
Fig.~\ref{fig_Q2_00a}.

The data for amplitude ratios were fitted by
\begin{equation}
\mathcal{Q} (L) = \mathcal{Q}_0 + b_i L^{y_i} + \cdots  \; ,
\label{fit_ratio}
\end{equation}
where exponent $y_i$ is left to be determined by numerical data.
The fits of the $Q(L)$ data yield $Q_0=0.4774 (4)$ and $y_i=-1.03(6)$,
and those of $Q_a$ give $Q_{a0}=0.6927 (4)$ and $y_i=-0.99(3)$. 
Both estimates of $y_i$ agree with integer $-1$, which can be 
explained as $y_{t2}-1$.

\subsection{$N=1.25, 1.50$, and $1.75$}

The analysis of the Monte Carlo data for the honeycomb $O(N)$ model
with $N=1.25, 1.50$, and $1.75$ follows an analogous procedure 
as that for $N=2$. Thus, the details of the fitting procedure are 
skipped here. The fitting results are listed in 
Tables~\ref{table_fit_energy} and ~\ref{table_fit_chi}.
\begin{table}[htb]
\centering
\begin{tabular}{rr|lllll}
\hline\hline
  $N$           &                & $\rho_l$          & $\rho_c$ 
& $n_b   $      & $C     $       & $C_a   $       \\
\hline \\[-2mm]
  $1.25$        & $y  $          & $0.887(2)$        & $0.884(5)$
& $0.885(5)$    & $0.885(4)$     & $1.378 (3)   $  \\[2mm]
                & $y  $(exact)   & $0.887\cdots $    & $0.887\cdots$
& $0.887\cdots$ & $0.887\cdots$  &   --            \\[2mm]
                & $r$            & $0.61297 (2) $    & $0.040207(2)$
& $1.09962(2)$  & $8.9   (1)$    & $2.7  (2)    $  \\[2mm]
\hline
  $1.5 $        & $y  $          & $0.745(3)$        & $0.74(1) $
& $0.75(2) $    & $0.74(1) $     & $1.406 (2)   $  \\[2mm]
                & $y  $(exact)   & $0.748\cdots $    & $0.748\cdots$
& $0.748\cdots$ & $0.748\cdots$  &   --            \\[2mm]
                & $r$            & $0.72952 (2) $    & $0.046245(2)$
& $1.13371(2)$  & $4.8   (2)$    & $1.21 (4)    $  \\[2mm]
\hline
  $1.75$        & $y  $          & $0.54(1)$         & $0.54(1) $
& $0.50(5) $    & $0.57(4) $     & $1.432 (3)   $  \\[2mm]
                & $y  $(exact)   & $0.554\cdots $    & $0.554\cdots$
& $0.554\cdots$ & $0.554\cdots$  &   --            \\[2mm]
                & $r$            & $0.86058 (1) $    & $0.051884(2)$
& $1.15514(2)$  & $3.160 (6)$    & $0.7  (9)    $  \\[2mm]
\hline
  $2.00$        & $y  $          & $0.1(2)$          & $0.004(8)$
& $-0.5(10)$    & $-0.01(3)$     & $1.5002(4)   $  \\[2mm]
                & $y  $(exact)   & $0     $          & $0          $
& $0         $  & $0        $    & $1.5         $  \\[2mm]
                & $r$            & $1.114835(4) $    & $0.057418(7)$
& $1.10956(2)$  & $1.7867(8)$    & $-0.46(3)    $  \\[2mm]
\hline\hline
\end{tabular}
\caption{\label{table_fit_energy}
Fitting results for energy-associated quantities. Symbol $r$ 
represents those analytical contributions, arising from the regular
part of the free energy. Also included are the predicted values for exponent
$y$ to be fitted. Symbol ``--" means that no prediction exists.}
\end{table}

\begin{table}[htb]
\centering
\begin{tabular}{rr|llllll}
\hline\hline
  $N$           &                & $\chi  $          & $\chi_a$ 
& $W_2   $      & $S_{2a}$       & $Q     $          & $Q_a   $ \\
\hline
  $1.25$        & $y  $          & $1.8327(2)$       & $1.8806(3)$
& $1.8804(3)$   & $1.9491(2)$    & $-1.6 (2)    $    & $-0.56(6)$    \\[2mm]
                & $y  $(exact)   & $1.8327\cdots$    & --           
& --            & --             & $-1.09\cdots $    & --             \\[2mm]
                & $\mathcal{Q}$  &                   &               
&               &                & $0.8069(2)   $    & $0.8245(7)$   \\[2mm]
\hline
  $1.5 $        & $y  $          & $1.7801(4)$       & $1.8843(2)$
& $1.8841(2)$   & $1.9499(1)$    & $-1.1 (2)    $    & $-0.60(4)$    \\[2mm]
                & $y  $(exact)   & $1.7805\cdots$    & --           
& --            & --             & $-1.09\cdots $    & --             \\[2mm]
                & $\mathcal{Q}$  &                   &               
&               &                & $0.7426(4)   $    & $0.7931(4)$   \\[2mm]
\hline
  $1.75$        & $y  $          & $1.7079(4)$       & $1.8854(2)$
& $1.8854(1)$   & $1.9497(1)$    & $-1.6 (2)    $    & $-0.79(8)$    \\[2mm]
                & $y  $(exact)   & $1.7078\cdots$    & --           
& --            & --             & $-1.46\cdots $    & --             \\[2mm]
                & $\mathcal{Q}$  &                   &               
&               &                & $0.6585(2)   $    & $0.7628(4)$   \\[2mm]
\hline
  $2.00$        & $y  $          & $1.5004(5)$       & $1.8751(1)$
& $1.8751(2)$   & $1.9444(1)$    & $-1.03(6)    $    & $-0.99(3)$    \\[2mm]
                & $y  $(exact)   & $1.5   $          & $1.875      $
& $1.875     $  & --             & $-1 ?        $    & $-1 ?    $     \\[2mm]
                & $\mathcal{Q}$  &                   &               
&               &                & $0.4774(4)   $    & $0.6927(4)$   \\[2mm]
\hline\hline
\end{tabular}
\caption{\label{table_fit_chi}
Fitting results for susceptibility-associated quantities. Symbol $\mathcal{Q}$ 
represents the universal values of those amplitude ratios at criticality.
We also give the predicted values for exponent
$y$ to be fitted. The exponent $y$ for amplitude ratios is 
for the dominant finite-size corrections. 
The question mark means that we are not very sure about the 
prediction of $y_i$. }
\end{table}

For $N=2$, quantities associated with those active loops or the spin 
variable $\sigma^{(1)}$ have clear physical meaning, since they appear
naturally in the $N=2$ IAT model described by Eq.~(\ref{def_H_AT}).
In this case, the scaling behavior of specific-heat-like quantity $C_a$ and
susceptibility $\chi_a$ can be predicted from the exact results for the
Baxter-Wu model or for the triangular IAT model. The numerical results fit
well with the predictions. Nevertheless, no exact results seem to exist for 
the finite-size behavior of the geometric quantity $S_{2a}$, the 
second-moment of the Ising clusters for spin $\sigma^{(1)}$.

For noninteger $N$, however, the symmetry between the active and the 
inactive loops breaks. 
To our knowledge, there are thus far no prediction for 
exponents governing the scaling behavior of quantities $C_a$ 
and $\chi_a$, as listed in Tables~\ref{table_fit_energy} and 
\ref{table_fit_chi}. On the other hand, the following scenario seems
plausible. The estimated exponent $y$ for $C_a$ in 
Table~\ref{table_fit_energy} is actually the mixed effect of $y_{t1}$
and $y_{t2}$, while that for $\chi_a$ is due to the effective mixture
of $y_{h1}$ and $y_{h2}$. We did try to make fits for the $C_a$
and $\chi_a$ by taking into account such a mixture effect,
but the results seem worse than those by Eqs.~(\ref{fit_specific_heat})
and (\ref{fit_chi}).

An observation is that, independent of the $N$ value, the exponents for 
the scaling behavior of $\chi_a$ and $W_2$ are alway equivalent. This 
strongly suggests that the percolation threshold of clusters formed
in the embedding SW step coincides with the thermal critical point
for any $N \geq 1$.  A rigorous proof is lacking.

\section{Dynamical behavior}

During the Monte Carlo simulations, the value of every 
sampled quantity was saved on hard disk after every sweep. 
Statistical analyses were then performed on these data. 
Both static and dynamic information of sampled quantities 
can then be obtained.

Let symbol $f$ be a given quantity, the unnormalized autocorrelation 
function is then defined as
\begin{equation}
C_f (t) \equiv \langle f(0) f(t) \rangle - \langle f \rangle ^2 \; .
\label{def_unnor_corr}
\end{equation}
The normalized autocorrelation function is 
\begin{equation}
\rho_f (t) \equiv C_f (t)/C_f(0) \; .
\label{def_nor_corr}
\end{equation}
Typically, $\rho (t)$ decays exponentially ($ \sim e^{-|t|/\tau} $) for large 
$t$, and the exponential autocorrelation time is defined 
as 
\begin{equation}
\tau_{exp,f} = \lim_{t \rightarrow \infty} \sup \frac{t}{-\ln |\rho_f (t)|} \; .
\label{def_tau_exp}
\end{equation}

In addition, we define the integrated autocorrelation time 
\begin{equation}
\tau_{int,f} = \frac{1}{2} \sum_{t = -\infty}^{\infty} \rho_f (t) = 
\frac{1}{2} + \sum_{t =1}^{\infty} \rho_f (t) \; .
\label{def_tau_int}
\end{equation}
Here, the factor $1/2$ is purely a matter of convention; 
it follows the definition in Ref.~\cite{Sokal_96}. 
The integrated autocorrelation time controls the statistical 
error in Monte Carlo measurements of $\langle f \rangle$. More precisely, 
the sample mean has variance 
\begin{equation}
\mbox{var} (\overline{f} ) \approx 
\frac{1}{n} (2 \tau_{inf,f}) C_f (0) \hspace{5mm} 
\mbox{for $n \gg \tau$} ,
\label{def_var}
\end{equation}
where $n$ is the total length of Monte Carlo simulations.

\subsection{$N=2$}

\begin{figure}
\begin{center}
\leavevmode
\epsfxsize 10.0cm
\epsfbox{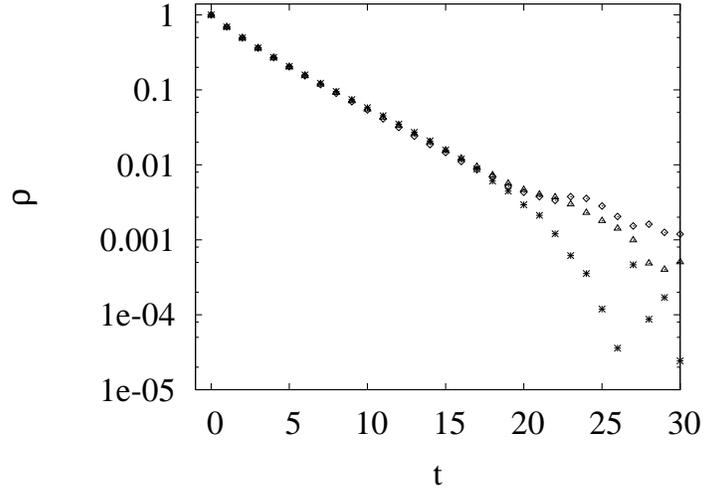}
\end{center}
\caption{Normalized autocorrelation function $\rho$ for susceptibility $\chi$.
The data points $\bigtriangleup$, $\Diamond$, and $\ast$ are for 
system sizes $L=160, 256, 360$, and $512$, respectively. The scattering 
behavior at the right-hand side is due to statistical noise. 
For $\rho_{\chi}\geq 0.01$, the $\rho$ data for different system sizes 
collapse rather well in this scale. This suggests that the critical slowing
down is absent in our embedding SW simulations of the $O(2)$ loop model on
the honeycomb lattice.}
\label{fig_pchi2_00a}
\end{figure}

\begin{figure}
\begin{center}
\leavevmode
\epsfxsize 10.0cm
\epsfbox{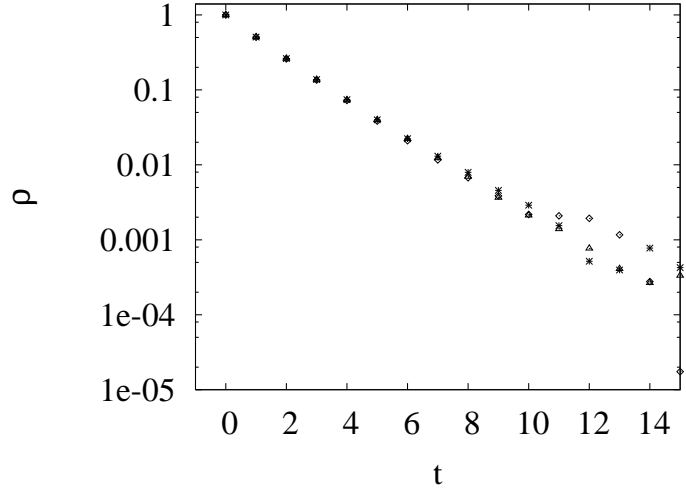}
\end{center}
\caption{Normalized autocorrelation function $\rho$ for the density of loop 
lengths $\rho_l$.
The data points $\bigtriangleup$, $\Diamond$, and $\ast$ are for
system sizes $L=160, 256, 360$, and $512$, respectively. The scattering
behavior at the right-hand side is due to statistical noise.
For $\rho_{\chi}\geq 0.01$, the $\rho$ data for different system sizes
collapse rather well in this scale. This suggests that the critical slowing
down is absent in our embedding SW simulations of the $O(2)$ loop model on
the honeycomb lattice.}
\label{fig_pS1a2_00a}
\end{figure}

\begin{figure}
\begin{center}
\leavevmode
\epsfxsize 10.0cm
\epsfbox{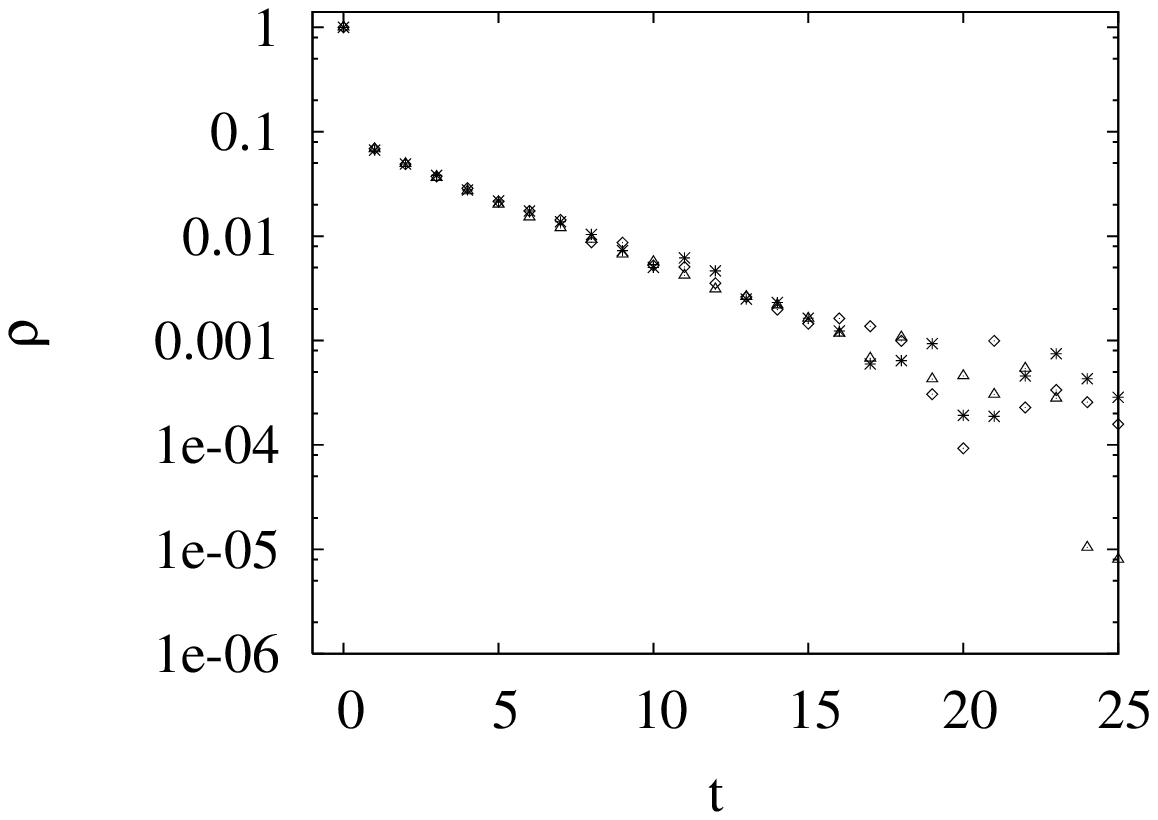}
\end{center}
\caption{Normalized autocorrelation function $\rho$ for susceptibility $\chi_a$
corresponding Isin spin variable $\sigma^{(1)}$.
The data points $\bigtriangleup$, $\Diamond$, and $\ast$ are for
system sizes $L=160, 256, 360$, and $512$, respectively. The scattering
behavior at the right-hand side is due to statistical noise.
For $\rho_{\chi}\geq 0.01$, the $\rho$ data for different system sizes
collapse rather well in this scale. This suggests that the critical slowing
down is absent in our embedding SW simulations of the $O(2)$ loop model on
the honeycomb lattice.}
\label{fig_pMi2_00a}
\end{figure}

Parts of the numerical data for the normalized autocorrelation 
function are shown 
in Fig~\ref{fig_pchi2_00a} for susceptibility $\chi$, 
in Fig.~\ref{fig_pS1a2_00a} the density of loop lengths $\rho_l$, 
and in Fig.~\ref{fig_pMi2_00a} for susceptibility $\chi_a$ for spin variable 
$\sigma^{(1)}$. 
The good collapse of the $\rho$ data strongly suggests that 
the critical slowing down is {\it absent } in the embedding SW simulation of 
the $O(2)$ loop model on the honeycomb lattice. 

For quantity $\chi_a$, 
after a single Monte Carlo step, the autocorrelation function drops to a value
smaller $0.1$. This means that two subsequent samples are almost effectively
independent. 

Except for very small $t$, the data lines in Figs.\ref{fig_pchi2_00a}, \ref{fig_pS1a2_00a},
and \ref{fig_pMi2_00a} are rather straight; the scattering phenomena at the 
right-hand side are due to statistical noise. Therefore, the autocorrelation function
is almost a purely exponential function of time: $\rho \approx e^{-t /\tau}$.

\begin{figure}
\begin{center}
\leavevmode
\epsfxsize 10.0cm
\epsfbox{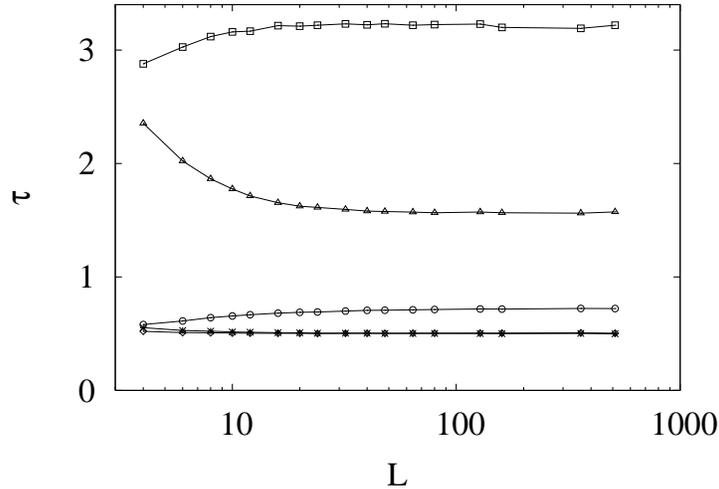}
\end{center}
\caption{Integrated autocorrelation time $\tau_{int}$ for $N=2$. 
The data points $\Box$, $\bigcirc$, $\bigtriangleup$, $\Diamond$, 
and $\ast$ are for susceptibility $\chi$, susceptibility $\chi_a$
for active spins, density of loop lengths $\rho_l$, the second 
moment $W_2$ of the clusters formed in Monte Carlo steps, and the 
bond-occupation density $n_b$, respectively. For the last two 
quantities, $\tau_{int}$ is very close to $0.5$, which means that 
autocorrelation function $\rho(t)$ is essentially zero for $t >0$. }
\label{fig_tau2_00a}
\end{figure}

The integrated autocorrelation time $\tau_{int}$ defined in Eq.~(\ref{def_tau_int}) 
was also measured for all sampled observables. Apart from the constant factor 
$1/2$, quantity $\tau_{int}$ is just the area underline the $\rho$ data line as
a function of time $t$. Parts of the $\tau_{int}$ data are shown 
in Fig.~\ref{fig_tau2_00a}.

We fitted the $\tau_{int}$ data for the sampled quantities by
\begin{equation}
\tau_{int} (L) = \tau_0 + L^{z}(b_1 +b_2 L^{-1} + \cdots) \; ,
\label{fit_tau}
\end{equation}
where the dynamic exponent $z$ is left free to be determined by the numerical
data. We assume that correction exponent $-1$, which appears in 
static quantities, also exists in the dynamical data. Nevertheless,
it turns out that, for $L \geq 8$, all the $\tau_{int}$ data for 
quantities in Fig.~\ref{fig_tau2_00a} can be fitted by Eq.~(\ref{fit_tau})
with $b_2=0$. The results are shown in Table~\ref{table_fit_tau}.
These show that, for all quantities shown in Table~\ref{table_fit_tau}, 
the integrated autocorrelation time  $\tau_{int}$ converge rapidly to 
constants as $L \rightarrow \infty$; the values of these 
constants are all rather small.
Quantity $\chi$ has the largest autocorrelation time $\tau_{int}$, and 
$\rho_l$ has the next largest $\tau_{int}$. 

\subsection{$N=1.25, 1.50$ and $1.75$ }

\begin{figure}
\begin{center}
\leavevmode
\epsfxsize 10.0cm
\epsfbox{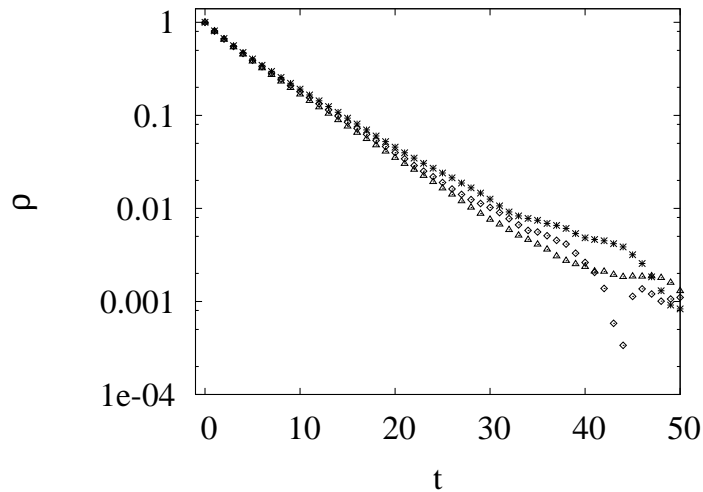}
\end{center}
\caption{Autocorrelation function $\rho_{\rho_l}$ for $N=1.25$.
The data points $\bigtriangleup$, $\Diamond$, and $\ast$ are for
system sizes $L=160, 256, 360$, and $512$, respectively. The scattering
behavior at the right-hand side is due to statistical noise. }
\label{fig_pS1a1_25}
\end{figure}

\begin{figure}
\begin{center}
\leavevmode
\epsfxsize 10.0cm
\epsfbox{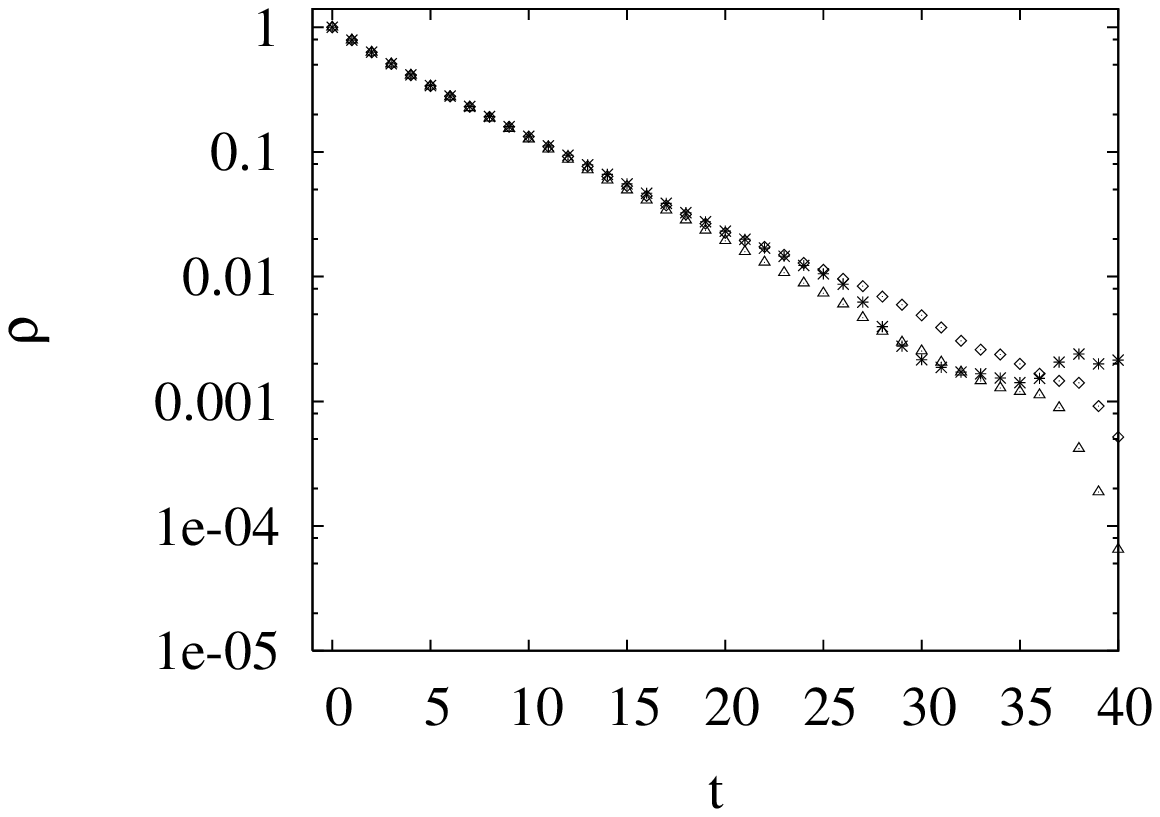}
\end{center}
\caption{Autocorrelation function $\rho_{\chi}$ for $N=1.50$.
The data points $\bigtriangleup$, $\Diamond$, and $\ast$ are for
system sizes $L=160, 256, 360$, and $512$, respectively. The scattering
behavior at the right-hand side is due to statistical noise. }
\label{fig_pMa1_50}
\end{figure}

\begin{figure}
\begin{center}
\leavevmode
\epsfxsize 10.0cm
\epsfbox{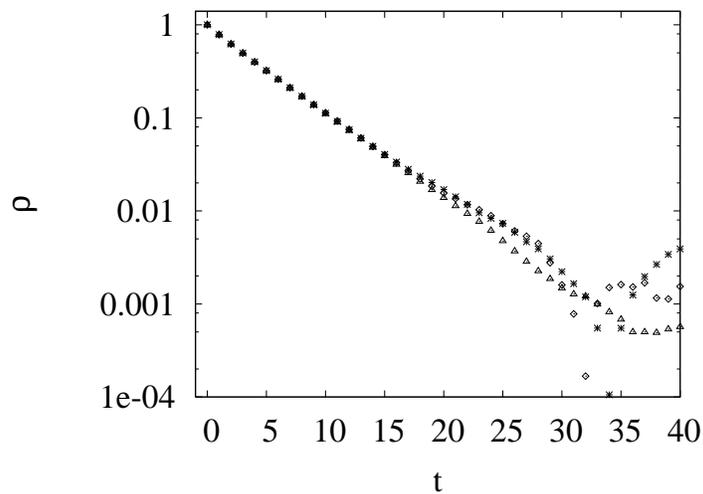}
\end{center}
\caption{Autocorrelation function $\rho_{\chi}$ for $N=1.75$.
The data points $\bigtriangleup$, $\Diamond$, and $\ast$ are for
system sizes $L=160, 256, 360$, and $512$, respectively. The scattering
behavior at the right-hand side is due to statistical noise. }
\label{fig_pMa1_75}
\end{figure}

For $N=1.25, 1.50$ and $1.75$, the general behavior of $\rho$ and $\tau$ 
for quantities listed in Table~\ref{table_fit_tau} is similar as that 
for $N=2$. Namely, the autocorrelation function $\rho (t)$ is almost a purely
exponential function of time $t$. For quantity $\chi_a$, after a single
Monte Carlo step, the value of $\rho$ drops to a significantly small value,
which is about $0.27$ for $N=1.25$. The quantities that have the largest 
two values of correlation time $\tau_{int}$ are $\chi$ and $\rho_l$; 
$\rho_l$ has the largest value for $N=1.25$ and $\chi$ for the others.
Therefore, we only show the $\rho_l$ data for $N=1.25$ in 
Fig.~\ref{fig_pS1a1_25} and the $\chi$ data for $N=1.50$
and $1.75$ in Figs.~\ref{fig_pMa1_50} and \ref{fig_pMa1_75}, respectively.

\begin{figure}
\begin{center}
\leavevmode
\epsfxsize 10.0cm
\epsfbox{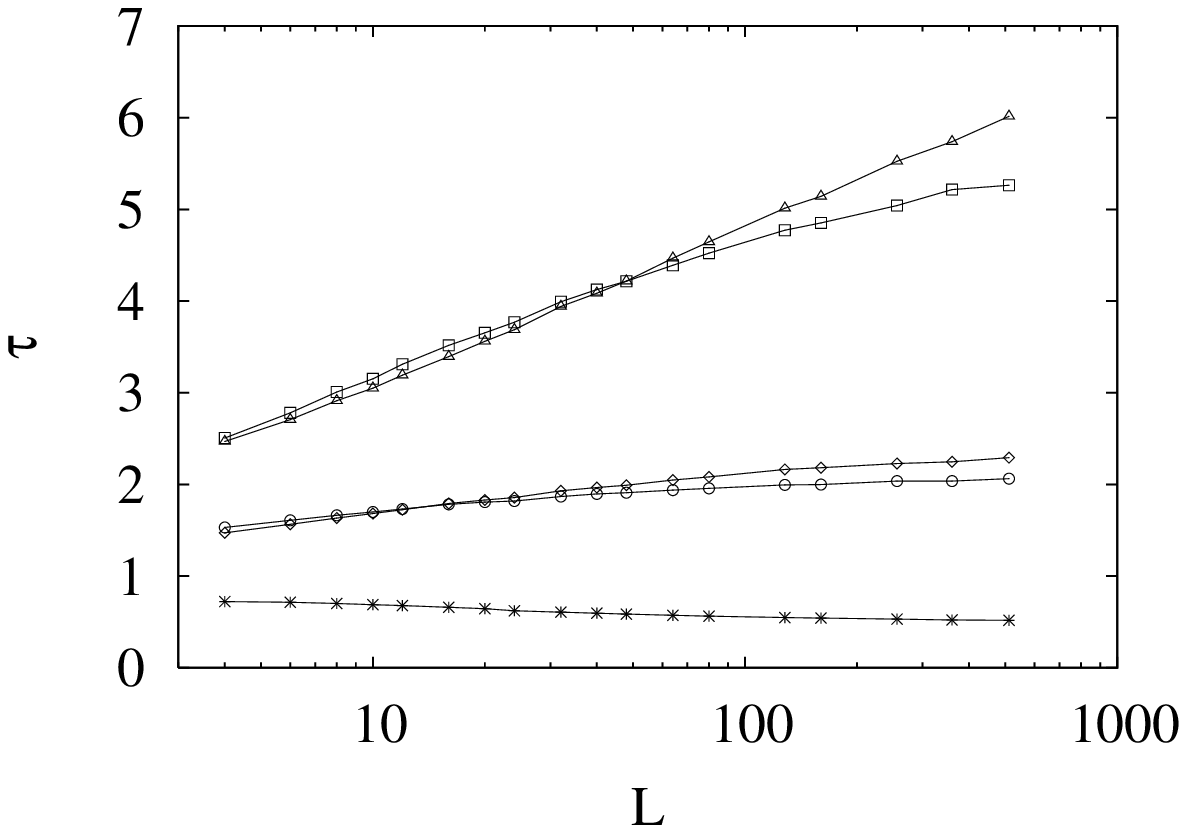}
\end{center}
\caption{Integrated autocorrelation time $\tau_{int}$ for $N=1.25$.
The data points $\Box$, $\bigcirc$, $\bigtriangleup$, $\Diamond$,
and $\ast$ are for $\chi$, $\chi_a$, $\rho_l$,  $W_2$ and $n_b$,
respectively.}
\label{fig_tau1_25}
\end{figure}

\begin{figure}
\begin{center}
\leavevmode
\epsfxsize 10.0cm
\epsfbox{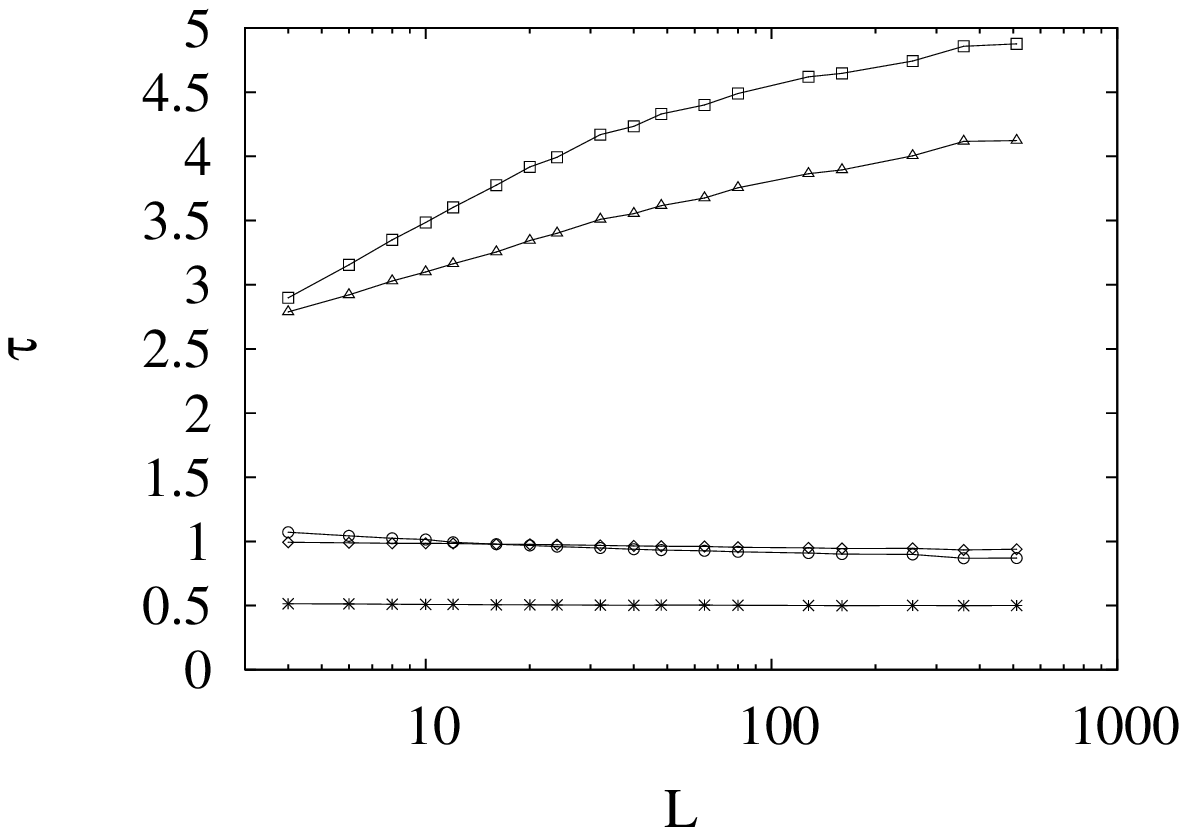}
\end{center}
\caption{Integrated autocorrelation time $\tau_{int}$ for $N=1.25$.
The data points $\Box$, $\bigcirc$, $\bigtriangleup$, $\Diamond$,
and $\ast$ are for $\chi$, $\chi_a$, $\rho_l$,  $W_2$ and $n_b$,
respectively.}
\label{fig_tau1_50}
\end{figure}

\begin{figure}
\begin{center}
\leavevmode
\epsfxsize 10.0cm
\epsfbox{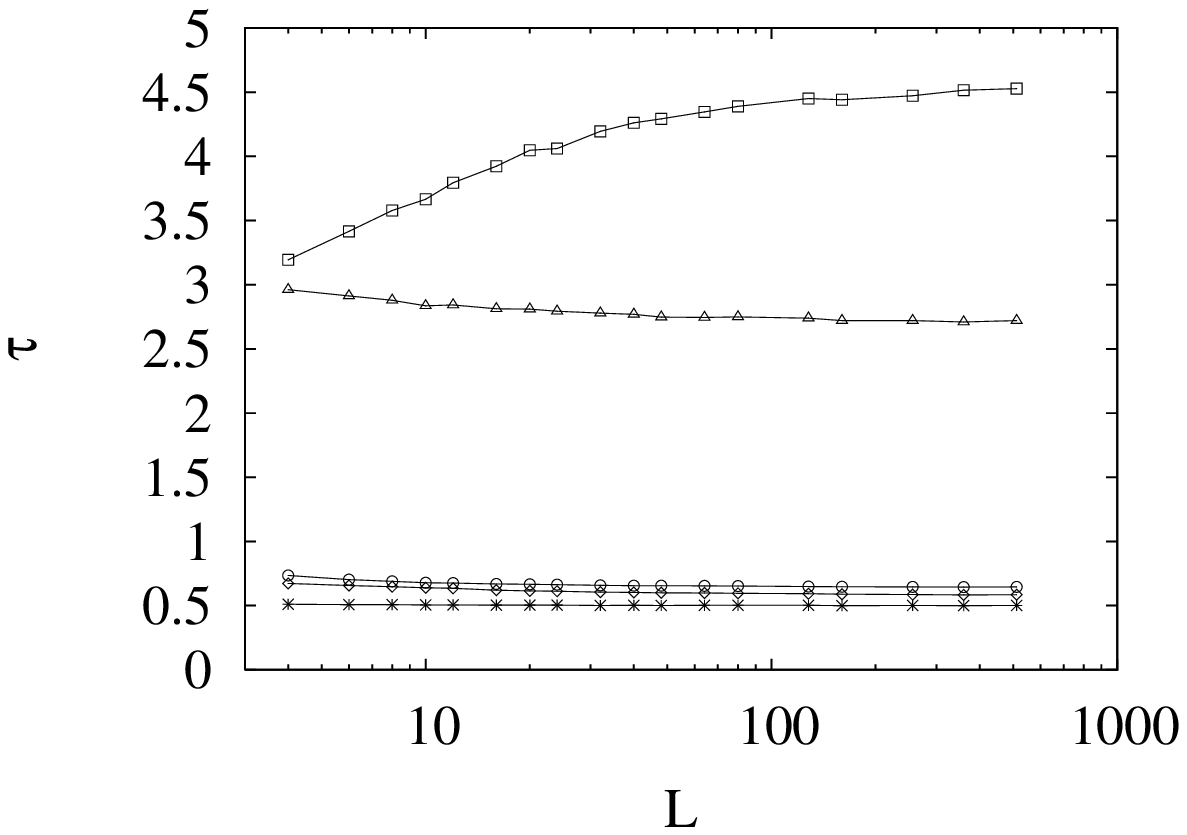}
\end{center}
\caption{Integrated autocorrelation time $\tau_{int}$ for $N=1.25$.
The data points $\Box$, $\bigcirc$, $\bigtriangleup$, $\Diamond$,
and $\ast$ are for $\chi$, $\chi_a$, $\rho_l$,  $W_2$ and $n_b$,
respectively.}
\label{fig_tau1_75}
\end{figure}

The $\tau_{int}$ data for the quantities in Table~\ref{table_fit_tau}
are also shown in Figs.~\ref{fig_tau1_25}, \ref{fig_tau1_50} and
\ref{fig_tau1_75} for $N=1.25, 1.50$, and $1.75$, respectively.

From Fig.~\ref{fig_tau1_25} for $N=1.25$, it is not completely clear whether
the critical slowing down is absent or not for quantities $\rho_l$ and 
$\chi$. Nevertheless, the approximately straight line of quantity $\rho_l$
suggests that, even the critical slowing down exists, the integrated correlation
time $\tau$ may diverge in a logarithmic format or in a power law of $L$
with a very small exponent $z$. 

We fitted the $\tau_{int}$ data by Eq.~(\ref{fit_tau}). 
For the $\tau_{int,\rho_l}$ data for $N=1.25$, it turns out 
they can still be described by Eq.~(\ref{fit_tau}) when a correction
term with $b_2$ is also included. 
For the other quantities and
for the other values $N$, the $\tau_{int}$ data for $L \geq 8$ all can 
be well described by Eq.~(\ref{fit_tau}) with $b=0$.
The results are shown in Table~\ref{table_fit_tau}; the value of $b_1$
is $ 2.5 (8)$ for $\tau_{int,\rho_l}$ with $N=1.25$. 

\begin{table}[htb]
\centering
\begin{tabular}{rr|llll}
\hline\hline
 Quantity      &                 & $N=1.25$         & $N=1.50$        
                                 & $N=1.75$         & $N=2.00$         \\
\hline \\[-2mm]
  $\rho_l$     & $\tau_0$        & $18(4)$          & $5.2(3)$
                                 & $2.70(2) $       & $1.566(3) $     \\  
               & $z     $        & $-0.06(2)$       & $-0.17(2)$
                                 & $-0.60(7)$       & $-1.7(1)  $     \\  
  $\chi  $     & $\tau_0$        & $7.7(2)$         & $5.18(10)$
                                 & $4.55(3)$        & $3.221(6) $     \\  
               & $z     $        & $-0.17(2)$       & $-0.43(3)$
                                 & $-0.80(7)$       & $-2.3(3)  $     \\  
  $\chi_a$     & $\tau_0$        & $2.18(2)$        & $0.82(3)$
                                 & $0.640(2)$       & $0.722(2) $     \\  
               & $z     $        & $-0.36(2)$       & $-0.28(4)$
                                 & $-0.64(8)$       & $-0.94(6)$     \\  
  $n_b   $     & $\tau_0$        & $0.499(4)$       & $0.498(4)$
                                 & $0.500(1)$       & $0.5002(4)$     \\  
               & $z     $        & $-0.65(3)$       & $-0.6(1)$
                                 & $-0.8(1)$        & $-1.2 (2)$     \\  
  $W_2   $     & $\tau_0$        & $2.7(1)  $       & $0.92(1)$
                                 & $0.577(3)$       & $0.5036(8)$     \\  
               & $z     $        & $-0.23(3)$       & $-0.42(4)$
                                 & $-0.54(6)$       & $-2.6 (8)$     \\  
\hline\hline
\end{tabular}
\caption{\label{table_fit_tau}
Fitting results for the integrated autocorrelation time $\tau_{int}$.}
\end{table}

Taking into account the approximate linearity of the $\tau_{int,\rho_l}$ 
data for $N=1.25$ in Fig.~\ref{fig_tau1_25}, we also fitted 
the $\tau_{int,\rho_l}$ data by 
\begin{equation}
\tau_{int}(L) = \tau_0 +\tau_i \ln L +b_1 L^{-1} \; .
\label{fit_tau_25}
\end{equation}
Indeed, all the data can be well described by Eq.~(\ref{fit_tau_25}); we 
obtain $\tau_0 =1.17(4)$, $\tau_i =0.79(2)$, and $b_1 =0.8(2)$.

\section{Discussion}

By making use of the equivalence of the loop configurations of the $O(N)$ loop
model and the low-temperature graph of the Ashkin-Teller model in the infinite-coupling
limit (IAT), we formulated an embedding Swendsen-Wang-type algorithm for the $O(N)$ 
loop model
with real value $N \geq 1$. For $N=1$, this algorithm reduces to the conventional
Swendsen-Wang method for the Ising model. With some modifications, an embedding 
Wolff-type method (single-cluster version) is readily available.

We then applied our cluster algorithm to the $O(N)$ loop model on the honeycomb lattice.
The numerical data reveal the finite-size scaling behavior of several quantities.
The associated exponents are confirmed to be those exact values predicted by
the Coulomb gas theory and by conformal field theory. 

The dynamical data strongly imply that the embedding cluster algorithm 
suffers little from critical slowing down. This is somewhat 
impressive in the sense that the dynamic exponent $z$ of 
the Swendsen-Wang type algorithm must satisfy 
the Li-Sokal bound~\cite{Li_89}:
$z \geq  \alpha /\nu$, unless it can be proved that the amplitude of 
terms with exponent $\alpha / \nu$ vanishes. The value of 
$\alpha/\nu \equiv 2 y_{t1}-2 $ can be easily calculated 
from Eq.~(\ref{critical_exponent}), which yields
$1.6654\cdots$, $1.5610\cdots$, $1.4156$, and $1$ for 
 for $N=1.25$, $1.50$, $1.75$, and $2$, respectively.
The absence of a dynamic exponent $z > \alpha \nu $ in the 
embedding SW simulations of the $O(N)$ loop model 
implies that the amplitude for terms with exponent $\alpha/\nu$ 
indeed vanishes.
We argue that this is because our embedding SW cluster algorithm has made
use of the symmetry among the $N$ colors of Ising spins.

Similar scenarios exist elsewhere. For instance, it can be proved that, 
in Metropolis simulations of the Ising model,  the dynamic exponent $z$ 
must satisfy $z \geq \gamma/\nu$. However, the dynamic exponent 
$z$ of the Swendsen-Wang algorithm is much smaller than $\gamma/\nu$. 
This is because the symmetry between the up- and down-pointing Ising
spins is fully taken into account in the Swendsen-Wang algorithm.

To have a better understanding of our above argument, let us consider 
another version of Swendsen-Wang-type cluster method for the IAT or 
the $O(N)$ model with integer $N$, as described in Sec. II. This algorithm 
directly simulates each color of the spin variables $\sigma^{(m)}$ in 
the IAT model for $N=1,2,\cdots,N $.  In the language of loop 
configurations, the algorithm only updates loops in the same color; it
does not interchange or reassign colors of loops. In other words,
the symmetry among loops in different colors is not taken into account.
Such a cluster algorithm has been applied to the critical $N=2$ IAT model on 
the triangular lattice, which is equivalent to the critical $O(2)$ model
on the honeycomb lattice or the Baxter-Wu model. It was found that 
the dynamic exponent $z$ for the integrated correlation time 
is $1.18(2)$; indeed, $z$ satisfies the Li-Sokal bound: $z > \alpha/\nu$.

Instead of the individual spin variables $\sigma^{(m)}$, the final version of 
the embedding SW algorithm simulates the product variable $s= \prod 
\sigma^{(m)}$. After each update of spin variable $s$, the colors 
of loops are reassigned randomly, irrespective of the existing colors
of loops. In this sense, it is natural that dynamic exponent $ z \geq 
\alpha/\nu$ vanishes.

One would then expect that the subleading exponent $\alpha'/\nu = 2y_{t2}-2$ 
serve as a lower bound for the dynamic exponent of our embedding SW cluster 
algorithm. 

The $N=1$ loop model is just the Ising model. From universality,
one has $\alpha'/\nu \equiv 2 y_{t2}-2 =0 $ for 
the $O(1)$ model on any planar graph. 
On the honeycomb lattice, Eqs.~(\ref{critical_exponent}) and (\ref{relation_gq})
tell that $\alpha'/\nu$ is a monotonically decreasing function 
of $q=N^2$. Therefore, one has $\alpha'/\nu <0$ for $N>1$.
In this sense, our cluster algorithm still 
satisfies the Li-Sokal bound. 
Since the value of $\alpha'/\nu$ decreases as a function of $N$, it is also 
expected that, as $N$ increases, the value of $\tau_{int}$ decrease. This is
consistent with $\tau_0$ in Table~\ref{table_fit_tau}.

It is clear from Sec. III that the embedding cluster algorithm described 
in this work cannot be applied to the $N <1$ case. Nevertheless, since 
all the loop configurations are the low-temperature graphs of the $s$ 
spin variable, and vice versa. It seems that, for $N < \approx 1$,
a reweighting modification of the Swendsen-Wang simulation of the Ising 
model can still be useful. Such a procedure can be described as
\begin{itemize}
\item {\bf Step 1.} For a spin configuration $\{ s \}$, in which the
number of loops is $c$, generate a new spin configuration $\{ s' \}$ by
using the Swendsen-Wang algorithm.
\item {\bf Step 2.} Derive the loop information for the new configuration
$\{ s' \}$, and calculate the loop number $c'$. Accept the new configuration
with probability $N^{c'-c}$, namely, set $s \leftarrow s'$. Otherwise, 
keep the old spin configuration $s$.
\end{itemize}
Repeating of these two steps forms a valid `cluster' algorithm 
for $N \leq 1$. For the $O(N)$ loop model on the honeycomb lattice, it turns out 
that, for $N \geq 0.8$ and small system sizes $L \leq 100$, this algorithm 
works pretty well.


\end{document}